# An Introduction to the Compute Express Link™ (CXL™) Interconnect

Debendra Das Sharma

Intel, debendra.das.sharma@intel.com

Robert Blankenship

Intel, robert.blankenship@intel.com

Daniel S. Berger

Microsoft Azure, daberg@microsoft.com

**Abstract.** The Compute Express Link (CXL) is an open industry-standard interconnect between processors and devices such as accelerators, memory buffers, smart network interfaces, persistent memory, and solid-state drives. CXL offers coherency and memory semantics with bandwidth that scales with PCIe bandwidth while achieving significantly lower latency than PCIe. All major CPU vendors, device vendors, and datacenter operators have adopted CXL as a common standard. This enables an inter-operable ecosystem that supports key computing use cases including highly efficient accelerators, server memory bandwidth and capacity expansion, multi-server resource pooling and sharing, and efficient peer-to-peer communication. This survey provides an introduction to CXL covering the standards CXL 1.0, CXL 2.0, and CXL 3.0. We further survey CXL implementations, discuss CXL's impact on the datacenter landscape, and future directions.

## 1 INTRODUCTION

The Compute Express Link (CXL) is an open industry standard that defines a family of interconnect protocols between CPUs and devices. While the CXL specification [3] and short summaries by news outlets are available, this tutorial seeks to provide technical details while staying accessible to a broad systems audience. The CXL protocol spans the entire compute stack and touches on many branches of computer science. To manage this breadth, we focus on fundamental engineering aspects, with less focus on security features, security implications, and higher-level software functions.

As a general device interconnect, CXL takes a broad definition of devices including graphics processing units (GPUs), general purpose graphics processing unit (GP-GPUs), field programmable gate arrays (FPGAs), as well as a wide range of purpose-built accelerators and storage devices. Traditionally, these devices use the Peripheral Component Interconnect-Express® (PCI-Express® or PCIe®) serial interface. CXL also targets memory which is traditionally connected to the CPU through the Double Data Rate (DDR) parallel interface.

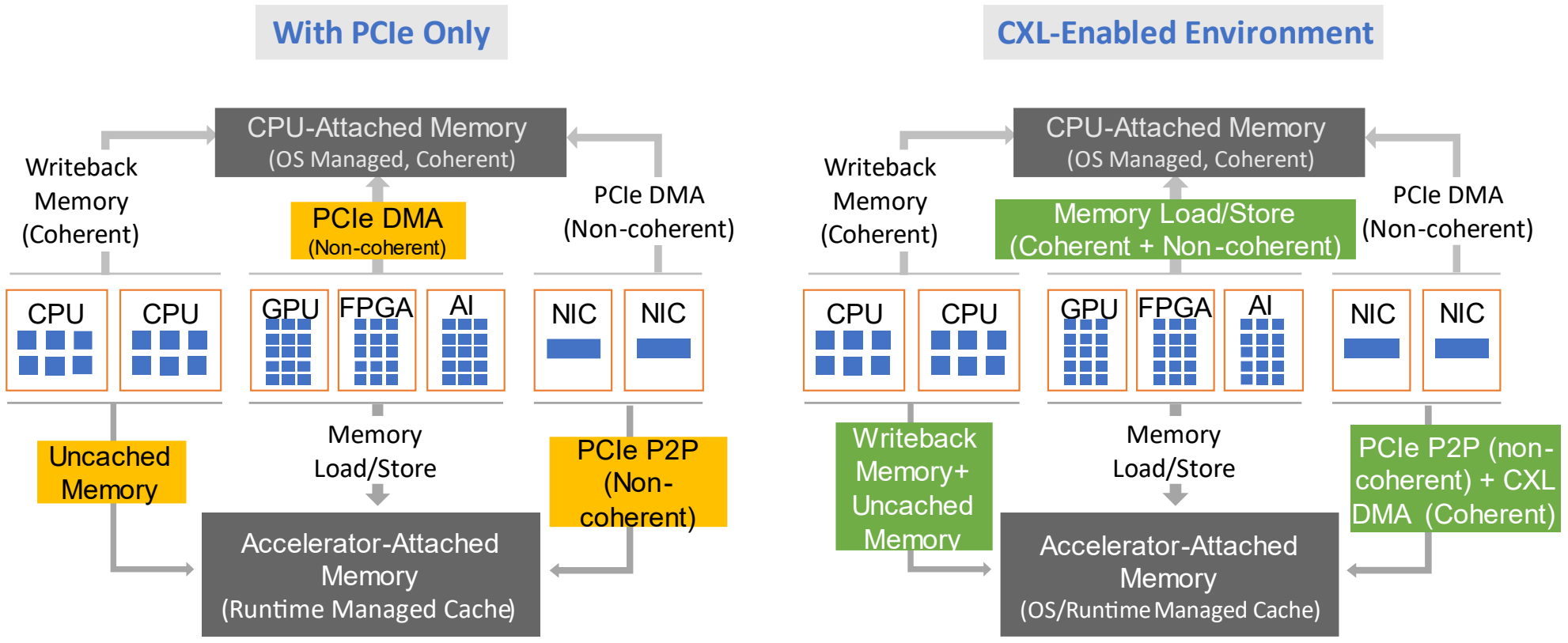

*Figure 1: CXL enables coherency and memory semantics and builds on top of PCIe' s physical subsystem.*

While PCIe and DDR have been great interfaces for a wide range of devices, they also come with some inherent limitations. These limitations lead to the following challenges that motivated the development and deployment of CXL.

**Challenge 1: coherent access to system and device memory.** System memory is conventionally attached via DDR and cacheable by the CPU cache hierarchy. In contrast, accesses from PCIe devices to system memory happen through non-coherent reads/writes, as shown in Figure 1. A PCIe device cannot cache system memory to exploit temporal or spatial locality or to perform atomic sequences of operations. Every read/write from the device to system memory passes through the host's root complex (RC) which keeps PCIe consistent with CPU caching semantics. Similarly, memory attached to a PCIe device is accessed non-coherently from the host, with each access handled by the PCIe device. Device memory thus cannot be mapped to the cacheable system address space.

Non-coherent accesses work well for streaming I/O operations (such as network access or storage access). For accelerators, entire data structures are moved from system memory to the accelerator for specific functions before being moved back to the main memory and software mechanisms are deployed to prevent simultaneous accesses between CPUs and accelerator(s). This poses a barrier for new usage models like Artificial Intelligence (AI), Machine Learning (ML), and smart network interface cards (NICs) where devices seek to access parts of the same data structures simultaneously with the CPU using device-local caches without moving entire data structures back and forth. This challenge also arises in the evolving area of processing-in-memory (PIM), which seeks to move computation close to data [80]. There is currently no standardized approach for PIM devices to coherently access data that may be present in the CPU cache hierarchy. This leads to cumbersome programming models which hinder the widespread adoption of PIM [67, 68, 80].

**Challenge 2: memory scalability.** Demand for memory capacity and bandwidth increases proportionate to the exponential growth of compute [17, 18, 29, 30, 33, 61, 62, 65]. Unfortunately, DDR memory has not been keeping up with this demand. This limits memory bandwidth per CPU. A key reason for this mismatch in scaling is the pin-inefficiency of the parallel DDR interface. Scaling up by adding DDR channels significantly adds to platform cost and introduces signal integrity challenges. In principle, PCIe pins would be a great alternative due to their superior memory bandwidth per pin, even with the added latency of serialization/ deserialization, as discussed later. For example, a x16 Gen5 PCIe port at 32 GT/s offers 256 GB/s with 64 signal pins. DDR5-6400 offers 50 GB/s with ~200 signal-pins. PCIe also supports longer reach with retimers[1], which would allow moving memory farther away from CPUs and using more than 15W of power per

1 Retimers enable extending the cable length between the host and a device beyond electrical limits defined in the specification (usually dozens of centimeters). They essentially retransmit a fresh copy of the signal, effectively doubling the channel length.

DIMM, resulting in superior performance. Unfortunately, PCIe does not support coherency and device-attached memory cannot be mapped to the coherent memory space. Thus, PCIe has not been able to replace DDR.

Another scaling challenge is that DRAM memory cost per bit has recently stayed flat. While there are multiple media types including Managed DRAM [71]. ReRam [72], 3DXP/Optane [73], the DDR standard relies on DRAM-specific commands for access and maintenance, which hinders adoption of new media types.

**Challenge 3: memory and compute inefficiency due to stranding.** Today's datacenters are inefficient due to stranded resources. A resource, such as memory, is stranded when idle capacity remains while another resource, such as compute, is fully used. The underlying cause is tight resources coupling where compute, memory, and I/O devices belong to only one server. As a result, each server needs to be overprovisioned with memory and accelerators to handle workloads with peak capacity demands. For example, a server that hosts an application that needs more memory (or accelerators) than available cannot borrow memory (or accelerators) from another underutilized server in the same rack and must suffer the performance consequences of page misses. On the other hand, servers where all cores are used by workloads often have memory remaining unused. Stranding has adverse power [36], cost [18], and sustainability [35] implications and has been the source of low resource utilization at Alibaba [29], AWS [34], Google [30], Meta [17, 33], and Microsoft [18,31,32].

**Challenge 4: fine-grained data sharing in distributed systems.** Distributed systems frequently rely on fine-grained synchronization. The underlying updates are often small and latency sensitive, as work blocks on updates. Examples include partition/aggregate design patterns in web-scale applications like web search, social network content composition, and advertisement selection [38, 39, 41, 42]. In these systems, query updates are often under 2kB (e.g., a search result). Other examples are distributed databases that rely on kB-scale pages and distributed consensus with even smaller updates [37, 44-46]. Sharing data at such fine granularity, means that the communication delay in typical datacenter networks dominates the wait time for updates and slows down these important use cases [38, 40]. For example, transmitting 4kB at 50GB/s (400Gbit/s) takes under 2us, but communication delays exceed 10us on current networks [40]. A coherent shared-memory implementation can help cut down communication delays to sub microseconds, as we will see later.

**The role of CXL**. CXL has been developed to address these four and other challenges. Since its first release in 2019, CXL has evolved through three generations (see Section 2). Each generation specifies the interconnect and multiple protocols (see Section 3) while remaining fully backward compatible. Table 1 overviews the three current CXL generations versions and key use case examples. The CXL 1.0 multiplexes coherency and memory semantics on top of the PCIe physical layer, as shown in Figure 1. CXL introduces custom link and transaction layers to achieve low latency comparable to remote socket memory accesses. This addresses both Challenge 1 (coherency) and Challenge 2 (memory scaling) by enabling CXL devices to cache system memory. This also standardizes a coherent interface to facilitate the broad adoption of PIM systems and programming models. CPUs can also cache device memory, which also addresses Challenge 4 (fine-grained distributed data sharing) for heterogeneous computing. Additionally, memory attached to a CXL device can be mapped to the system cacheable memory space. This facilitates heterogeneous processing and helps with memory bandwidth and capacity expansion challenges demonstrated in **Error! Reference source not found.**. CXL 1.0 also continues support for PCIe's non-coherent producer-consumer semantics [1,2,8].

CXL 2.0 additionally addresses Challenge 3 (resource stranding) by enabling resource pooling across multiple hosts. We use host to refer to a single-socket or multi-socket system under the control of a single operating system or hypervisor. Pooling overcomes resource stranding and fragmentation by reassigning resources (e.g., memory) to different hosts over time, without having to reboot these hosts. The CXL protocol enables pooling by introducing CXL switches that build a small network of hosts and memory devices.

CXL 3.0 addresses Challenge 3 on a larger scale with multiple levels of CXL switching. This enables building dynamically composable systems at the rack or even the pod level. Furthermore, CXL 3.0 addresses Challenge 4 (distributed data sharing) by enabling fine-grained memory sharing across host boundaries.

CXL requires CPU and device support. The widespread adoption of CXL into commercial products by virtually all silicon vendors is a testament to the technology gaining widespread traction due to its ability to solve real-world problems.

This puts CXL on a viable path to solve key industry challenges with broad deployment [8,15,16,17,18]. CXL 3.0 has also reached a level of maturity that facilitates reviewing its fundamental design choices. This tutorial introduces background in Section 2. We detail incremental releases of CXL 1.0, CXL 2.0, and CXL 3.0 in Sections 3, 4, and 5, respectively. Section 6 discusses CXL implementations and performance. Section 7 discusses broader impacts and future directions.

*Table 1 Overview of CXL specification generations, speeds, and use cases*

| Generation | Year | Scope | Speed | Use case |
|---|---|---|---|---|
| CXL 1.0 & CXL 1.1 | 2019 | Single machine | 32 GT/s | Accelerators (Challenge 1)<br>Bandwidth and capacity expansion (Challenge 2) |
| CXL 2.0 | 2020 | 2-16 machines (single switch) | 32 GT/s | Small-scale resource **pooling** (Challenge 3) |
| CXL 3.0 | 2022 | 100s machines (multiple switches) | 64 GT/s | Large-scale resource pooling and **sharing** (Challenges 3 and 4) |

## 2 CXL BACKGROUND AND DESIGN CHOICES

Cache-coherent interconnects have historically been ‘symmetric’, as they focused on multi-socket systems or multi-GPU systems. Every node (CPU or GPU) would implement the same coherency algorithms and track required state. Common examples are Intel’s Quick Path Interconnect (QPI) and Ultra Path Interconnect (UPI), AMD’s Infinity Fabric, IBM’s Bluelink, and Nvidia’s NVLink. Over the years, multiple extensions of symmetric interconnects sought to address some of the challenges of Section 1, e.g., CCIX and OpenCAPI. Several companies also licensed protocols like QPI. However, industry experience showed that complexity and performance demands of symmetric cache coherence pose significant deployment challenges as the mechanisms to orchestrate coherency varies widely across different architectures and design choices, which vary over time. Additionally, not all use cases require coherency support, e.g., Challenge 2 (memory scaling) and Challenge 3 (stranding).

Wide-spread adoption also calls for a plug-and-play ecosystem that is compatible with previous generations and across different system architectures. Backwards compatibility has proven to be one of the main reasons why PCIe is so successful as a ubiquitous I/O interconnect for more than two decades. It facilitates technology transitions independently for device and platform manufacturers. For example, a platform may migrate to PCIe Gen 5 whereas the SSD vendors may decide to stay with Gen 4 and migrate at a later point of time based on the technology evolution of SSDs. It also helps protect customer investments as a customer may reuse some older generation device(s) in a new platform, e.g., to reduce carbon emissions [35].

At Intel, the idea of adding simplified coherency mechanisms on top of PCIe goes back to 2005, motivated by accelerators needing to cache system memory. This resulted in transaction processing hints to use the CPU’s cache hierarchy for faster device accesses and atomics semantics being added to PCIe 3.0, while keeping PCIe non-coherent. Intel also initially pursued adding memory semantics on top of PCIe 3.0 to enable pooling with a shared memory controller (SMC). However, PCIe 3.0 bandwidth, 8.0 GT/s, severely limited the number of servers that could pool resources with one SMC. Multi-SMC topologies would have resulted in higher latency due to multiple hops. Another challenge was cabling. In 2019, PCIe 5.0 at 32.0 GT/s (and progress on PCIe 6.0 at 64.0 GT/s) and strong cable support revived efforts at Intel. This resulted in Intel Accelerator Link (IAL): a proprietary protocol with both caching and memory support on PCIe 5.0. Leveraging the experience from developing PCI-Express over two decades, Intel donated the IAL 1.0 specification and launched the CXL consortium with Alibaba, Cisco, Dell, Google, Huawei, Meta, Microsoft, and HPE in March 2019. IAL 1.0 specification was renamed as CXL 1.0 specification.

CXL adopts an *asymmetric approach to coherency*, *backwards-compatibility*, and *openness* to enable a diverse and open ecosystem that facilitates broad deployment. CXL coherence is decoupled from host-specific coherence protocol details. The host processor is also responsible for orchestrating cache coherency for simplicity of implementing coherency in devices. A device’s caching agent enforces a simple MESI (modified, exclusive, shared, invalid) coherency protocol with a small command set. CXL supports multiple use cases by offering multiple protocols that differ in complexity and devices can implement only a subset of protocols (Section 3).

CXL utilizes the PCIe physical layer and devices plug into PCIe slots. The backward compatible evolution of CXL (like PCIe) as well as its interoperability with PCIe ensures that companies can make their investment in CXL with guaranteed interoperability with prior-generation CXL devices as well as any PCIe device. Building on PCIe infrastructure lowers the barrier to entry by enabling reuse of IP building blocks, channels, and software infrastructure. From a SoC viewpoint, a multi-protocol capable PCIe physical layer, as shown in Figure 2, helps reduce silicon area, pin count, and power [1, 2, 8, 9].

When CXL was launched in 2019, the industry was fragmented with competing interconnect standards such as OpenCAPI, GenZ, and CCIX. Since then, CXL membership has grown to about 250 companies with all CPU, GPU, FPGA, networking, IP providers actively contributing within the consortium. After Intel donated CXL 1.0 in March 2019, the consortium published CXL 1.1 adding compliance test mechanisms in September 2019. Following that, the consortium published CXL 2.0 and CXL 3.0 in November 2020 and August 2022, respectively, including more usage models, while maintaining full backward compatibility. Over time, the industry has coalesced around CXL. For example, competing standards GenZ and OpenCAPI have donated their IP and funds to CXL to rally behind a common standard.

## 3 CXL 1.1 PROTOCOL

CXL's first generation introduces coherency and memory semantics for devices directly attached to a host. This enables fine-grained heterogeneous processing of shared data structures for CPUs and accelerators (Challenge 1 in Section 1) as well as cost-effective scale-up for memory bandwidth and capacity (Challenge 2 in Section 1).

CXL is an asymmetric protocol, like PCIe. The host processor contains the 'Root Complex' (RC), one per CXL link, each connected to a device which is an 'End Point'. The host processor orchestrates cache coherency, as described below. Software configures the system through instructions executing in the host processor, which generates the configuration transactions to access each device. CXL supports x16, x8, and x4 link widths natively and x2 and x1 widths in degraded mode. Degraded mode refers to a PCIe link automatically going into a narrower width and/or lower frequency to overcome higher than expected error rates on a given lane. CXL supports data rates of 32.0 GT/s and 64.0 GT/s natively while 16.0 GT/s and 8.0 GT/s data rates are supported in degraded mode [1, 5, 8].

Figure 2 illustrates how CXL offers full interoperability with PCIe since it uses the PCIe stack. A CXL device starts link training in the PCIe Gen 1 Data Rate of 2.5 GT/s and negotiates CXL as the operating protocol using the alternate protocol negotiation mechanism defined in the PCIe 5.0 and PCIe 6.0 specifications if its link partner can support CXL.

### 3.1 CXL Protocols and Device Types

CXL is implemented using three protocols, CXL.io, CXL.cache, and CXL.memory (aka. CXL.mem), which are dynamically multiplexed on PCIe physical layer. Figure 2 illustrates CXL's multi-protocol support. CXL.io protocol is based on PCIe. It is used for device discovery, status reporting, virtual to physical address translation, and direct memory access (DMA). CXL.io uses the non-coherent load-store semantics of PCIe [7, 8, 9]. Existing PCIe software infrastructure will be reused with the necessary enhancements to take advantage of the new capabilities such as CXL.cache and CXL.mem. CXL.cache is used by a device to cache system memory. CXL.mem enables CPUs and other CXL devices to access device memory as cacheable memory. CXL.mem makes it possible for memory attached to a device to be cacheable (referred to as 'Host managed Device Memory' – HDM), similar to the host memory, resulting in a host's uniform view across HDM and host memory (Figure 2). CXL.io is mandatory for all devices whereas CXL.cache and CXL.mem are optional and usage specific, as shown in Figure 3.

Figure 3 shows three device types defined in CXL 1.0/1.1 specification, that capture various usage models. Type 1 devices are accelerators such as smart NICs that use coherency semantics along with PCIe-style DMA transfers. Thus, they implement only the CXL.io and CXL.cache protocols. Type 2 devices are accelerators such as GP-GPUs and FPGAs with local memory that can be mapped in part to the cacheable system memory. These devices also cache system memory for processing. Thus, they implement CXL.io, CXL.cache and CXL.mem protocols. Type-3 devices are used for memory bandwidth and capacity expansion and can be used to connect to different memory types, including supporting multiple memory tiers attached to the device. Thus, Type-3 devices would implement only the CXL.io and CXL.mem protocols. CXL Type-3 devices offer a cost, power, and pin-efficient alternative to adding more DDR channels to server CPUs, while offering flexibility in system topologies due to longer trace lengths which results in alleviating power delivery as well as cooling constraints [8,9].

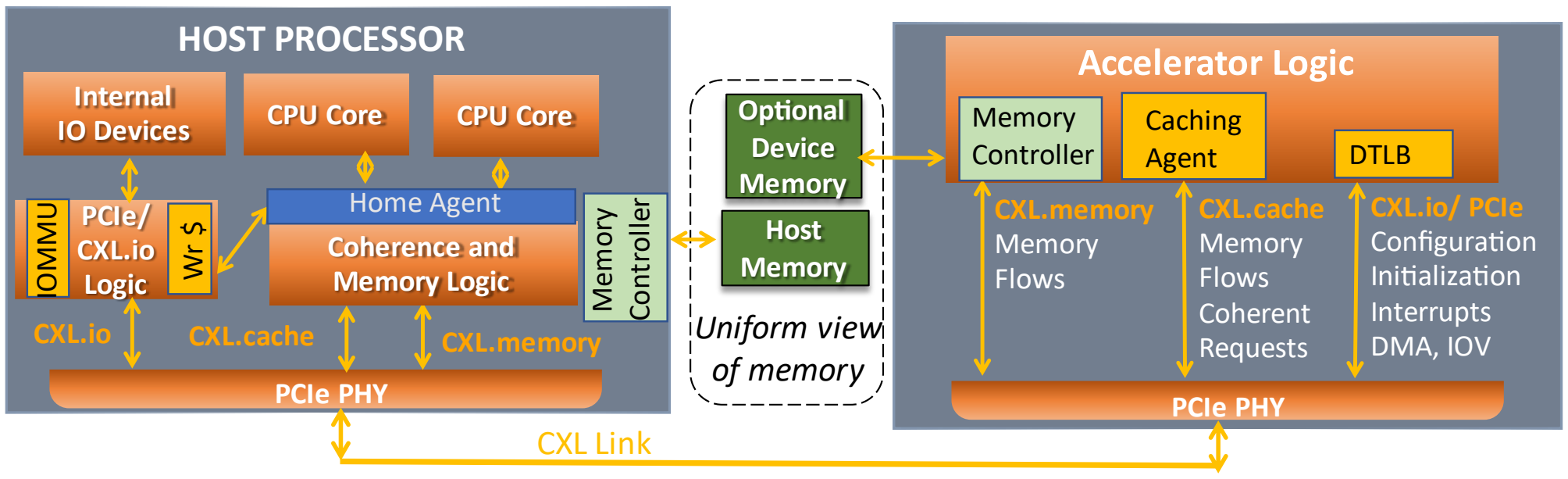

*Figure 2: Dynamic multiplexing of three protocols on PCIe physical layer with CXL [8]*

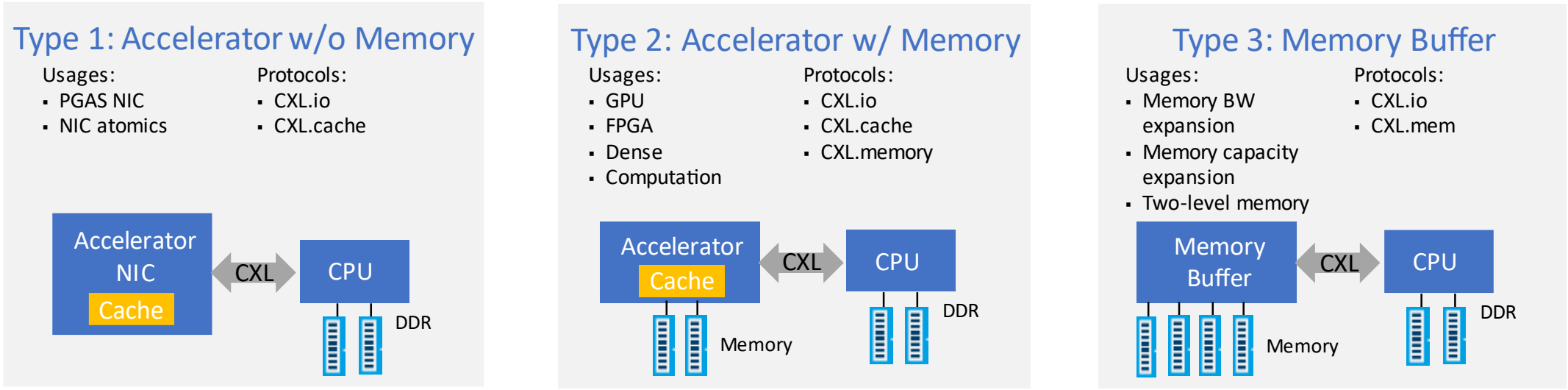

*Figure 3: Three broad categories of CXL end-point devices [8]*

CXL adopts a layered protocol approach. The physical layer is responsible for physical information exchange, interface initialization, and maintenance. The data link layer (or link layer) is responsible for reliable data transport services and establishing a logical connection between devices. The transaction layer(s) handles the transactions associated with each protocol along with any architectural ordering semantics, flow control, and credits. Each of these layers has an architected set of registers that software accesses for configuring, controlling, and obtaining status of the link. Readers are referred to the respective specifications [5, 7] for details.

### 3.2 CXL 68-Byte Flit

CXL multiplexes its three protocols at the PCIe physical layer. The unit of transfer for each protocol is a Flit (Flow-Control Unit). CXL 1.0, 1.1, and 2.0 specifications define a 68-byte Flit [1, 2, 3, 4, 8, 9]. CXL 3.0 specification introduces additional 256-byte Flits [5, 6, 7, 8, 9, 13, 14] that are discussed in Section 5. Each 68-byte Flit comprises of a 2-byte protocol-ID, a 64-byte payload, and a 2-byte CRC (Cyclic Redundancy Check) protecting the payload (Figure 4a). The protocol-ID delineates different types of packets and has built-in redundancy to detect and correct multiple bit flips within those 2 bytes. The CRC guarantees detection of up to 4 random bit flips in the 64-byte payload (plus the 2-bytes of CRC), which results in very low (<< $10^{-3}$) failure in time (FIT, which is the number of failures in a billion hours in the link) for the specification mandated bit error rate (BER) of less than $10^{-12}$ [5, 7]. Figure 4b shows the 16-byte slot mechanism used by CXL.cache, CXL.mem, and Arb/Mux Link Management Packets (ALMP), where 4 16-byte slots make up the 64-byte payload of the Flit. The Arb/Mux sends ALMP packets to its link partner's Arb/Mux unit to coordinate link management functionality such as power management across the multiple stacks as needed. Figure 4c shows the layout for the all-data

Flit used by CXL.cache and CXL.mem protocols, referred to as CXL.cache+mem, since the disambiguation between the two occurs in the link layer as they both are natively Flit based. CXL.io uses the PCIe transaction/ data-link layer packets (TLP/ DLLP) as-is which are sent in the payload part of the Flit. Since the TLP and DLLP have their own CRC (32bits and 16bits respectively) [7], the 16-bit Flit CRC is ignored by the receiver for CXL.io packets [2,3].

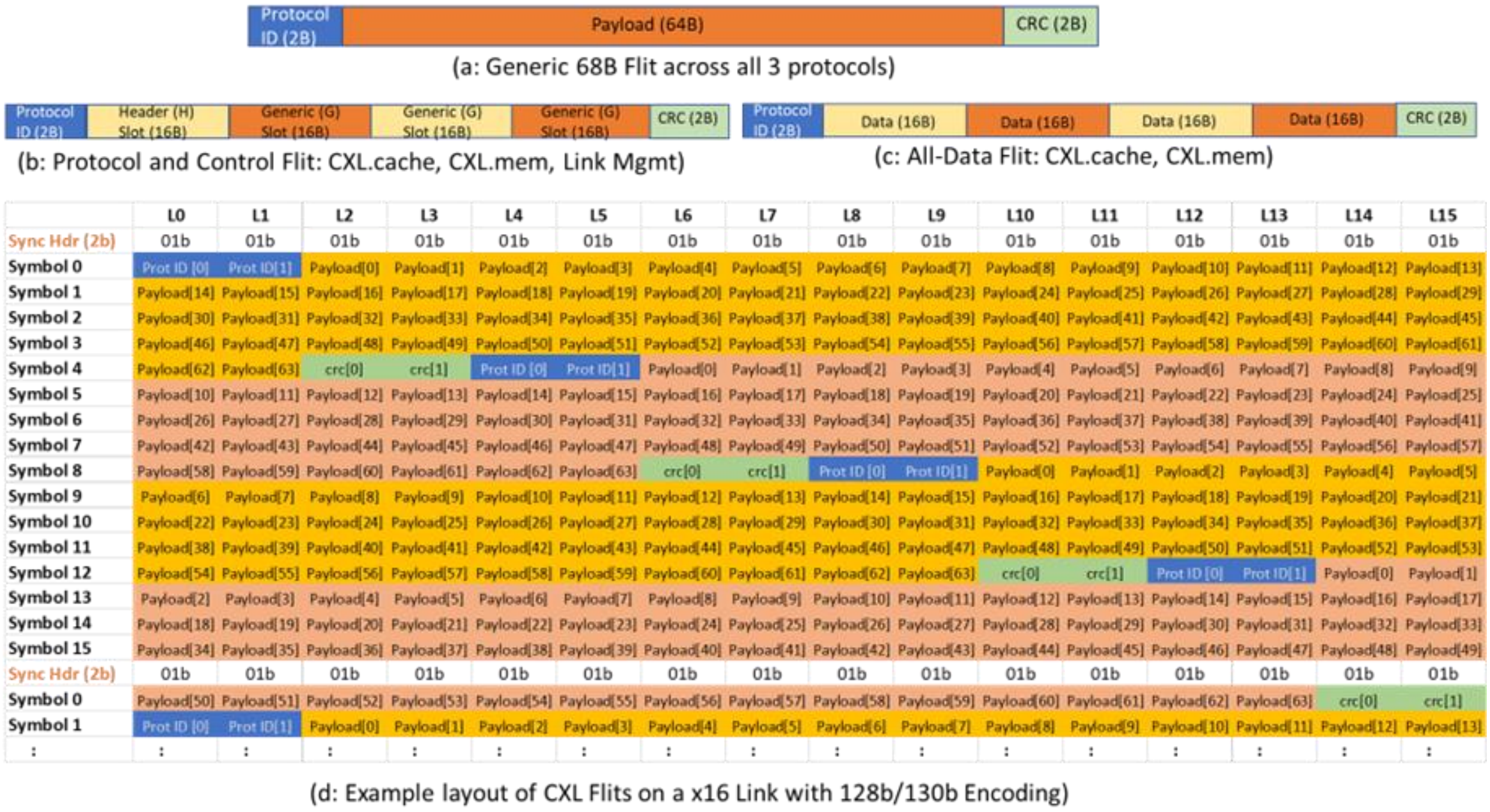

*Figure 4: CXL 68-Byte Flit Layout [2,8]*

CXL.cache and CXL.mem accesses have low-latency, similar to a native CPU-to-CPU symmetric coherency link [1, 8, 9]. Thus, memory access latency from a CXL device would be similar to memory access from a DDR bus in a remote socket. While this is higher than memory access from a DDR bus in a local socket [81], in a 2-socket symmetric multi-processing system, it is acceptable due to NUMA (non-uniform memory access) optimization and the higher bandwidth resulting in lower latency in non-idle systems [8,9,10]. The Link layer and Transaction layer paths for CXL.cache and CXL.mem have low latency since they are natively Flit based and the choice of 64-byte payload of a Flit is identical to the cache-line sized transfers of these protocols. Muxing at the PHY level (vs higher level of the stack) helps deliver a low latency path for CXL.cache and CXL.mem traffic. This eliminates the higher latency in the link and transaction layers of PCIe/ CXL.io path due to their support for variable packet size, ordering rules, access rights checks, etc. conforming to the fundamental guiding principle of CXL specifications [1, 3]. The physical layer disambiguates between CXL.io, CXL.cache-mem, ALMP, and NULL Flits (when nothing is sent). Each of these 4 types of Flits have two cases to indicate whether the current Flit is the end of the data stream (EDS) prior to sending Ordered Sets. These Ordered Sets are injected by the physical layer for functionality such as periodic clock compensation or any link recovery event. These 8 encodings use 8-bits with a guaranteed hamming distance of 4 which are repeated twice to enable correction as well as detection [2,3]. Each Ordered Set is one Block (130b) long and is used for link training as well as clock compensation. Each Lane independently sends its Ordered Set(s).

While operating at 32 GT/s or lower, CXL uses the 128b/130b encoding scheme of PCIe where a 2-bit sync header is prepended for every 128bits of data on each Lane to distinguish between Data blocks vs. Ordered Set blocks [7,8]. The 128-bit data payload on each Lane is used to transmit Flits, as shown in the two data blocks of a x16 link in Figure 4d. A Flit can straddle across multiple data blocks and a data block can contain multiple Flits. Since CXL.io packets (TLP and DLLP) are not natively Flit based, a TLP or DLLP can straddle multiple Flits and a Flit may contain up to two packets.

Characterizing these overheads leads to the payload-throughputs discussed in Section 6.3. For latency optimization, CXL enables removal of the 2-bit sync header when all components of the link (including Retimers, if any) advertise support for this optimization during the initial PCIe link training process while negotiating CXL protocol [2].

### 3.3 CXL.io Protocol

The CXL.io protocol is based on PCIe and is used for functions such as device discovery, configuration, initialization, I/O virtualization, and DMA using non-coherent load-store semantics. CXL.io, like PCIe, is a split-transaction, credit-based, packetized protocol [7]. Split transaction means that any completion(s) associated with a CXL.io request transaction arrive(s) independently and asynchronously at a later time. This is a departure from early PCI-bus based architecture, where the transaction locks up the bus till its completion is provided. This is possible because each transaction is sent as an independent Transaction Layer Packet (TLP) and consumes 'credits' consistent with the buffer space it will consume on the Receiver. There are three flow-control classes (FCs): posted (P, for memory writes and messages), non-posted (NP, those that require completion such as memory reads and configuration/ I/O read/writes), and completions (C). CXL.io follows PCIe Ordering rules across these FCs, as summarized and explained in Figure 5 (CXL 3.0 introduces an exception to these ordering rules, as described in Section 5). These rules ensure forward progress while enforcing the producer-consumer ordering model. CXL.io mandates two virtual channels for ensuring quality of service (QoS). Traffic with different requirements and latency characteristics are put in different virtual channels (VCs) to minimize the effect of top of queue blocking in the system. For example, bulk traffic for bandwidth vs. latency-sensitive isochronous traffic can be put in different VCs and the arbitration policies can be set differently. Similarly, one may put DRAM accesses and persistent memory accesses in two different VCs since they have very different latency and bandwidth characteristics to ensure that traffic to one does not impact the other.

| | Posted (P) (Col 1) | Non-Posted (NP) (Col 2) | Completion (C) (Col 3) |
|---|---|---|---|
| P (e.g., Mem Wr) (Row A) | a. No<br>b. Y/N | Yes | a. Y/N<br>b. Yes |
| NP (e.g., Mem Rd) (Row B) | a. No<br>b. Y/N | Y/N | Y/N |
| C (with/out data) (Row C) | a. No<br>b. Y/N | Yes | a. Y/N<br>b. No |

The entry 'No' ensures producer-consumer ordering. For example, a Memory Write should not bypass a prior memory Write transaction (A1a), as the latter may indicate the Flag signifying that the prior write (Data) completed. The Flag bypassing the Data indicates someone may read stale Data.

The entry `Yes' enforces forward progress. For example, completions must be able to bypass prior non-posted (C2) since non-posted generates
Completion. So completions being stuck behind a prior Memory Read will deadlock the system.

For performance optimization, the Y/N applies where the ordering rules are relaxed. For example, a posted or completion with relaxed ordering (RO) attribute is permitted to pass another posted request (A1b, B1b, C1b)

*Figure 5: PCI-Express/ CXL.io Ordering Table for CXL 1.1 where 'Yes' indicates that a transaction (2nd Trans) can overtake a prior transaction (1st Trans) to ensure forward progress. To ensure producer-consumer ordering, 'No' indicates that overtaking is not allowed. For performance optimization, 'Y/N' applies where ordering rules are relaxed.*

Figure 6a describes the producer-consumer ordering model which acts as the contract between hardware and software. This is enforced across the entire system even for transactions that cross different hierarchies and for address locations that are in different memory locations and may have different attributes (e.g., cacheable vs. non-cacheable). The device may be the CPU, a CXL/PCIe device, or any entity within a switch. The basic idea is that the producer produces (memory write) data and subsequently writes the flag; the data and flag can be in multiple memory locations across the system. If the consumer sees the flag, it can then read the data and be assured that it obtains the latest data written by the producer. The flag can be in a ring buffer in system memory, a memory-mapped I/O location in a device, or an interrupt sent to the processor. The ordering rules delineated in Figure 5 ensure the producer-consumer ordering model. Another effect of the ordering rules is device synchronization usage, as represented in Figure 6b. Here the two devices may be executing two tasks and use A and B as indicators that they completed the tasks. If the read obtains old data, the process in that device can be suspended and be rescheduled by the device that completes the execution later. If (a, b) were possible, both processes in the two devices would be suspended for ever, an undesired outcome.

While CXL.io/PCIe ordering model can be used for some types of synchronization, the synchronization usage represented in Figure 6c cannot be enforced with pipelined accesses, since writes can bypass prior reads. This limitation causes smart NICs with partitioned global address space (PGAS) to serialize writes after reads when ordering matters. CXL.cache overcomes this limitation; the device can prefetch all data out of order and complete the transactions with low latency in the local cache conforming to the program order.

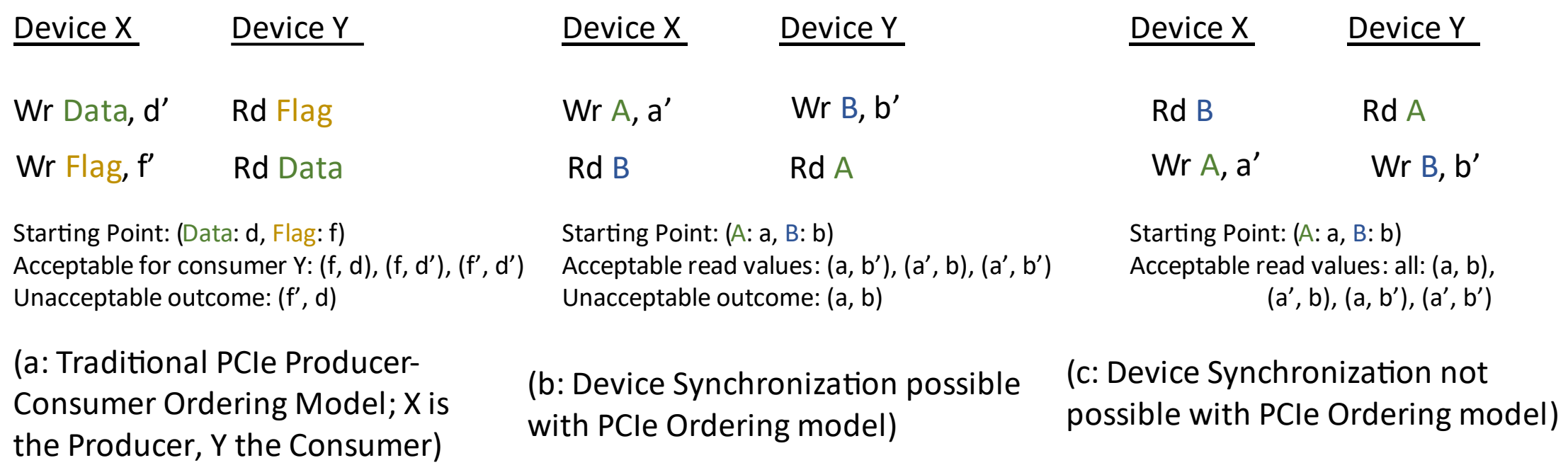

*Figure 6: Producer-Consumer ordering model in CXL.io and PCIe*

The CXL.cache+mem is mostly unordered but has some ordering constraints on a per cache line basis as will be discussed later. These constraints can be worked around in a topology that supports multiple paths between a source-destination pair by having a routing mechanism that ensures the transactions with dependencies for any given cache line always follow the same path. However, with the traditional PCIe (and CXL.io) the ordering requirements are across the entire memory space. Hence traditional PCIe (and hence CXL) must follow the tree topology. With CXL 3, we relax the CXL.io ordering requirements and any fabric topology can be supported, as described in Section 5.

CXL.io uses the standard PCIe DLLPs for exchanging information such as credits, reliable TLP delivery, power management etc. [2, 3, 7]. CXL.io uses the configuration space of PCIe and enhances it for CXL usage. This helps with using the existing device discovery mechanism. We expect the PCIe device driver would make the necessary enhancements to take advantage of the new capabilities such as CXL.cache and CXL.mem and the system software will program the new set of registers associated with the new capabilities. This builds on the existing infrastructure and makes the adoption of CXL easy for the ecosystem.

### 3.4 CXL.cache Protocol

The CXL.cache protocol enables a device to cache host memory using the MESI coherence protocol [49] with a 64-byte cache line size. To keep the protocol simple at the device the host manages all tracking of coherence for peer caches and the device never directly interacts with any peer cache. The protocol is built on 3 channels in each direction. The direction of the channels are Host-to-Device (H2D) and Device-to-Host (D2H). Each direction has a Request, Response, and Data channel. The channels flow independently in each direction with one exception: a Snoop message from the host in H2D Request channel must push a prior Global Observation (GO) message in the H2D Response for the same cache line address. The GO message indicates the coherence state (MESI) and the coherency commitment point to the device.

CXL.cache uses only host physical addresses. Caching devices, like CPUs, operate on virtual addresses when they execute code. Similarly, a high-performing device is expected to work with virtual addresses to avoid a software layer to provide the translation. The device is expected to implement a Device Translation Look-aside Buffer (DTLB), similar to CPU's TLB, to cache page table entries. It uses the Address Translation Services (ATS) of PCIe (and CXL.io) [2, 3, 5, 7] to fetch the virtual to physical translation as well as access control for ensuring proper isolation between multiple VMs/containers. Additionally, CXL extends ATS to convey if access to the address is allowed to use CXL.cache or is limited to only CXL.io. Since the DTLB is non-coherent, ATS expects the host processor to track the entries pending in DTLBs across the platform and initiate invalidation to those DTLB entries (or entry) that software invalidates. The host processor completes the invalidation handshake with the DTLB(s) using its ATS hardware mechanism prior to indicating completion to software that is invalidating page table entries.

The D2H Request channel includes 15 commands broken into 4 categories: Read, Read0, Read0-Write, Write. The Read category allows for a device to request coherence state and data for a cache line as shown in Figure 7. The response will be a coherence state in the H2D Response channel (if coherence is required) and a H2D Data. The Read0 category is a request just for coherence state with '0' data (meaning no data is required). This may be used to upgrade existing data in the cache (S to E) or to bring in E-state when the entire cache line is expected to be written. The H2D Response channel indicates coherence state provided by the host. The Read0-Write category allows the device to write data to the host directly without having any coherence state prior to issue. The H2D Response will indicate 'WritePull' when the host is ready to accept the D2H Data and the host will resolve coherence before indicating when that data is 'Globally Observed' (GO) by the system. The Write category is used for the device to evict data from the device cache. These requests can be for dirty data (M-state) or for Clean Data (E or S state). The host will indicate 'WritePull' for cases where device needs to provide data and it will indicate GO.

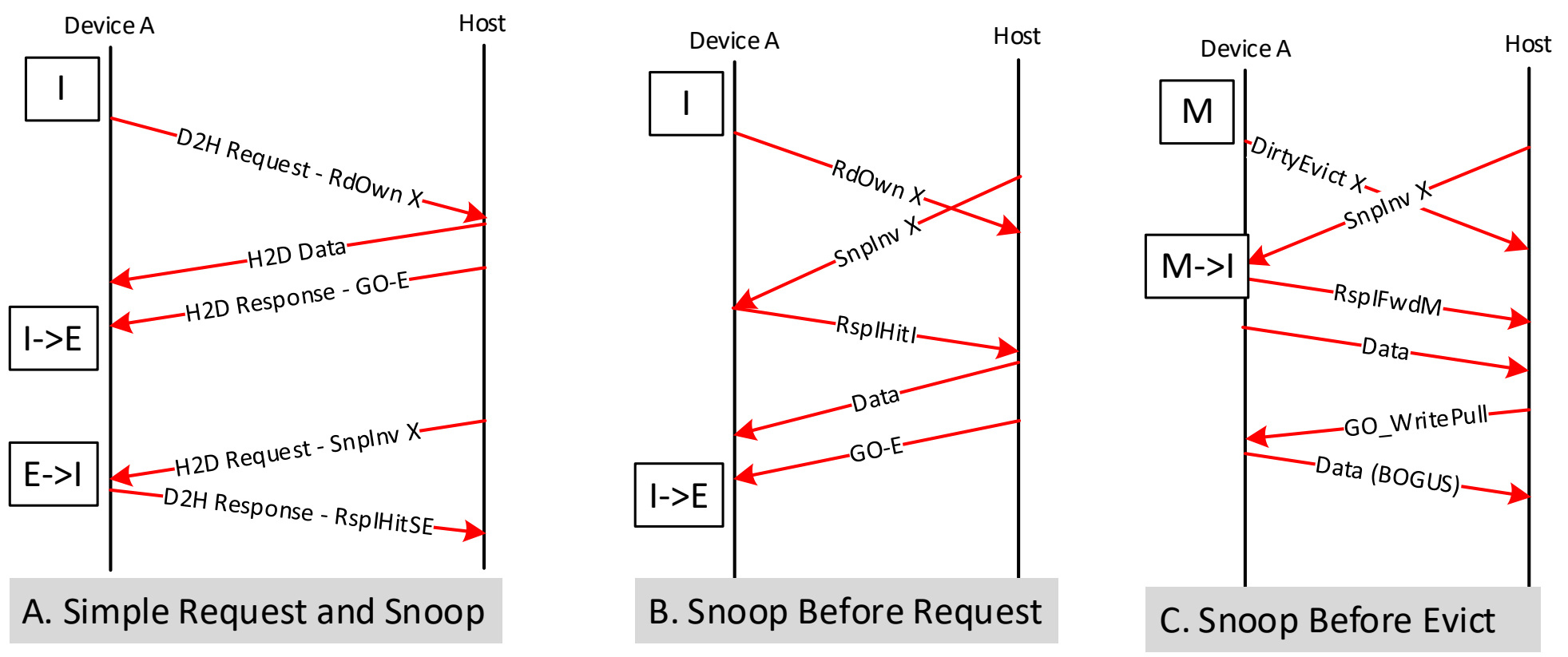

*Figure 7 CXL 1.1 Cache Protocol Flows*

The H2D Request channel is used for the host to change coherence state in the device which is referred to as 'Snooping' (abbreviated Snp). The device must update its cache as required by the snoop type and in the case of cache with dirty data (M-state) it must also return that data to the host. Figure 7 provides an example where the host sends a SnpInv X which requires the device to invalidate the cache for Address X. The device changes cache state from E to I for Address X and then sends a RspIHitSE on the D2H Response channel. The RspI meaning its final cache state is I-state and HitSE indicating the cache had S or E-state prior to downgrading to final I-state.

A critical component of a coherence protocol is how it handles conflicting accesses for the same address. Two cases are important to highlight: (1) Req-to-Snoop, and (2) Eviction-to-Snoop. For Req-to-Snoop for the device determines how to respond, the GO message on H2D Response channel must be ordered in front of any future Snoop to the same cache line address. The reason for this is illustrated in Figure 7A where the GO message sent from the host must be observed before the later snoop so the device is aware of that it has Exclusive ownership of the address and can process the snoop correctly. Figure 7B is the reverse order case where the host is processing the snoop first and will stall future requests until the snoop is completed, so the device observing snoop before GO results in processing the snoop while the cache is in Invalid state. For Evict-to-Snoop Figure 7C shows the case where snoop arrives while a DirtyEvict is outstanding, and device must reply to the snoop with the current M-state data. The later GO_WritePull must still return the data but includes indication that the data is bogus (meaning may be stale) so the host must drop the data as newer data may exist in other agents.

### 3.5 CXL.mem Protocol

The Memory Protocol enables a device to expose Host-managed Device Memory (HDM) which allows the host to manage and access this memory similar to native DDR connected to the host. The protocol is independent of the media used using a simple set of reads and writes with Host Physical Addresses[2] requiring the device to translate internally into the device's media address space. There is limit to the amount of CXL-attached memory inherent in the protocol; in practice limitations come from the number and size of attachable CXL devices. For accelerator devices that wish to also cache this memory, advanced semantics are provided to enable the device to directly cache the memory, relying on the device to track host caching of this memory.

The protocol uses two channels in each direction denoted as Master-to-Subordinate (M2S) and Subordinate-to-Master (S2M). The M2S direction provides a Request channel and a Request-with-Data (RwD) channel. The S2M direction provides a Non-Data-Response (NDR) channel and Data-ReSponse (DRS) channel. To allow simple/low latency implementations there is no ordering between channels.

The two use cases for CXL.mem are: (1) as 'host memory expander', (2) as accelerator memory exposed to the host. Both use the HDM term to describe the memory region accessed by the CXL.mem protocol, but their protocol requirements differ. To identify these requirements a suffice indicates where a host memory expander is 'host-only coherent' with a '-H' modifier (HDM-H) or the accelerator memory exposed to the host is 'Device-managed coherent' with a '-D' modifier (HDM-D)[3]. HDM-H does not include any coherence protocol assumptions, whereas HDM-D includes cache state and cache snooping attributes used in each message. HDM-D also requires a method for communication of desired cache state of the memory from the device to host referred to as the 'Bias Flip' Flow. [4]

HDM-H optionally provides 2-bits of Meta Value with each cache line for host use. Possible uses are coherence directory, security attributes, or data compression attributes. Figure 8A shows an example flow with a host read that is updating the Meta Value in the device and the device returning the prior state of the Meta Value. In this example the device is required to change the current Meta Value stored in the device as the value changed from 2 (old value) to 0 (new value). Note that the Memory Media storage of Meta Value and ECC (Error Correcting Code) bits are device specific.

---

2 This contrasts with traditional DRAM technology interfaces like DDR4 which may rely on special commands for access and maintenance like Refresh or Bank Pre-charge before Read or write access and use local Device specific address space.

3 The original CXL 1.0 and 1.l specification did not use the -D and -H term, but instead the attribute was implied based on device type where a Type-3 device was HDM-H, and Type-2 device was HDM-D. This alignment to device types is removed in CXL 3.0 because of new requirements, so this paper's descriptions will be using CXL 3.0 terms to improve consistency.

4 "Bias Flip" flow in reference to the device Bias-Table (or coherence directory) which tracks if the host could have a cached copy of the address. The Bias tracking may indicate Host-S (Shared only in the host), Host-A (Any cache state in the host), or Device (The host does not have a cached copy of the address so referred to as "device bias"). The "Flip" is in reference to the Bias-Table changing the state being tracked from "Host Bias" to "Device Bias".

HDM-D uses the Meta Value field differently such that it exposes host coherence state to the device in this field, which allows the device to know what state the host is caching for each address in the HDM-D region. The specification uses the term Device Coherence (DCOH) agent to describe the agent in the device managing/tracking coherence between the device and the host. Figure 8B shows a case of HDM-D where the host is reading the data to cache in S-state and it requires the device to check its cache (Dev $) for a current copy, but in this example the cache does not have the data, so the data is delivered from host memory.

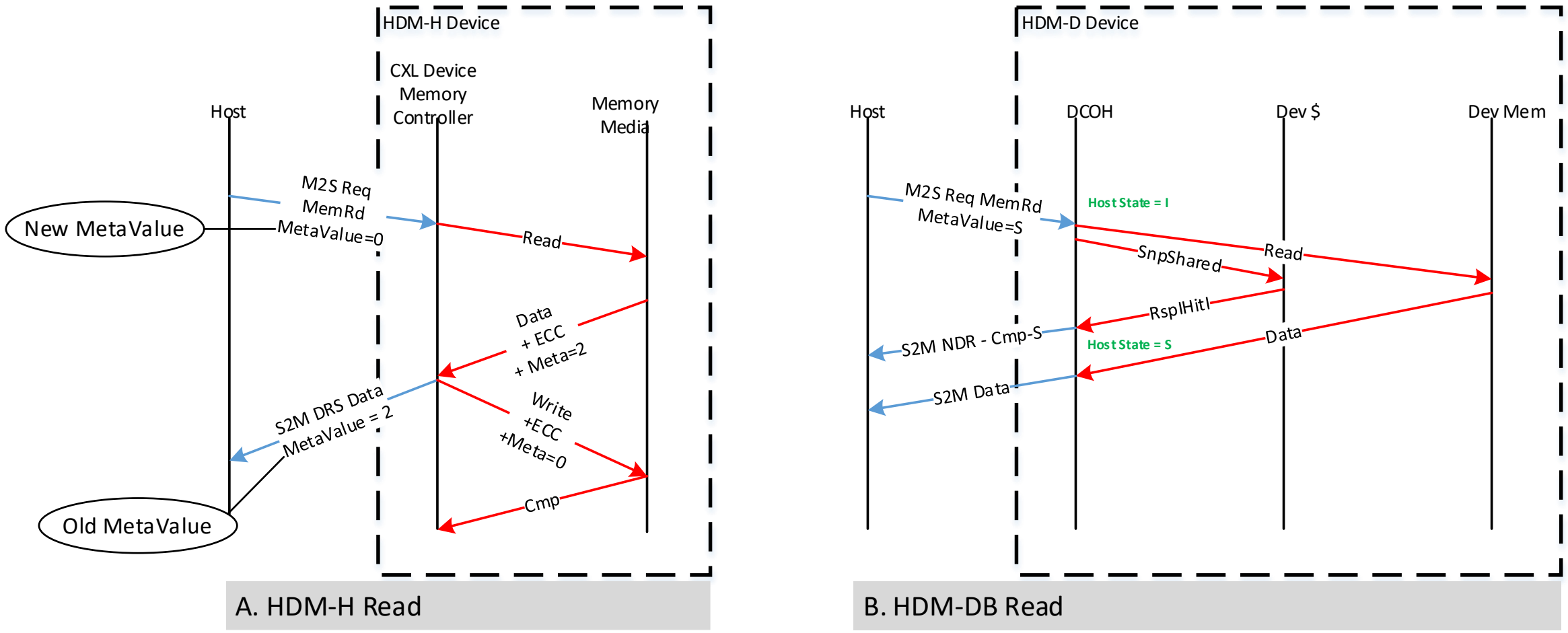

*Figure 8 CXL 1.1 Memory Protocol Flow*

The HDM-D coherence model allows the device to change the state of the host using a CXL.cache request. CXL.cache is the natural way to do this as HDM-D is defined only for Type-2. The flow, where the device changes host cache state, is referred to as 'Bias Flip' flow. An example of this flow is shown in Figure 9.The device sends 'RdOwnNoData X' to the host. The host detects the address X as an HDM-D address owned by the issuing device and will behave differently than it would have otherwise in that it will directly change cache state in the host forcing all host caches to I-state. The response to a 'Bias Flip' flow is different from traditional CXL.cache responses, in that the host will send response MemRdFwd message in this example on the CXL.mem M2S Req channel to indicate the host has completed the cache state change. By using the CXL.mem M2S Req channel it avoids race conditions with the host for future CXL.mem requests to the same address requiring the M2S Request channel to be ordered for access to the same cache line. The ordering requirement applies only to HDM-D addresses.

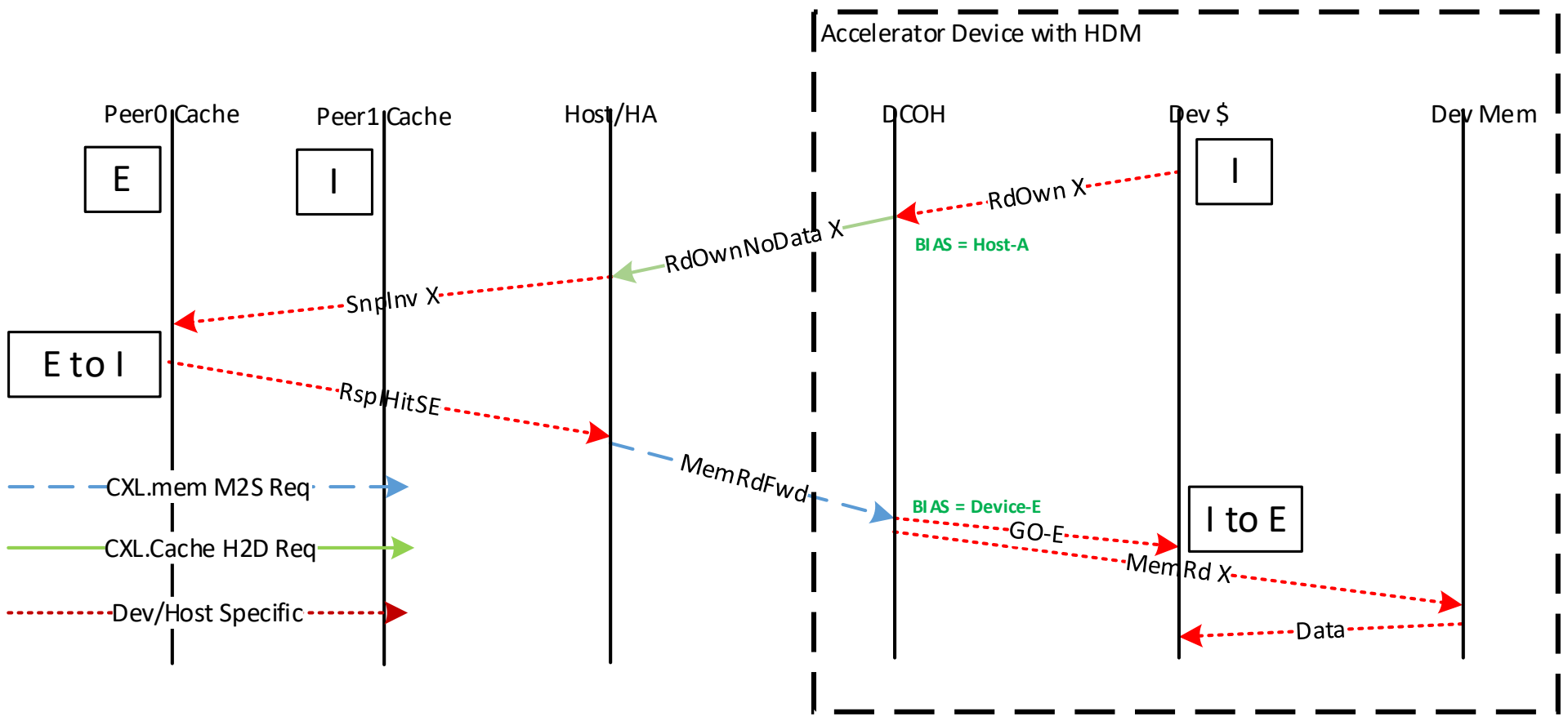

*Figure 9 HDM-D Bias Flip Flow*

### 3.6 Protocol Hierarchy

CXL is compatible with a range of coherence implementations inside the CPU. Figure 10-A captures a possible host CPU where CXL.cache is under the host L3 cache that is shared with CPU cores. In this example, the CXL.io/PCIe protocol access to host memory also passes through the same cache following destination coherent semantics that are used on the path to host memory. This diagram shows a single L3 in the socket. Alternatively, host architecture may split up L3 or other internal caches without impact on CXL.cache which is unaware of host cache details. The host scales to multiple CPU sockets or caches as defined in the host specific protocol (Proprietary CPU-to-CPU) which defines an internal Home Agent to resolve coherence between host caches (which incorporate CXL.cache devices). This host Home Agent may include host-specific optimizations for resolving coherence such as on-die snoop filters or in-memory directory states. The CXL.mem protocol is behind the Host Home Agent logic and can support simple memory expansion that is host only coherent (HDM-H) or device coherent (HDM-D). With HDM-D, an additional level of coherence resolution is included after the host Home Agent which allows the device to be the final arbiter of coherence for addresses owned by the device.

Figure 10-B shows the protocol dependence graph expected in CXL. The protocol dependence graph is defining which protocol channels may be dependent (or blocked) by other protocol channels. The dependence graph is a method to show deadlock freedom between channels if no circular dependence is created. An example dependence from the diagram would be that L1 Req may depend on L1-Snp channel to complete before a new L1 Req may be processed. This dependence exists because the request may require a snoop to be sent and completed before the request can be completed. The L1 protocol is CXL.cache showing an abstraction of the channels provided in CXL.cache where L1-Req maps to D2H Req, L1-Snp maps to H2D-Req and L1 RSP maps to H2D/D2H RSP & Data channels which are pre-allocated to drain into host or device, enabling combining these two for the dependence graph. The L2 protocol is host specific and shown as an example of channels that may exist for a host, but other channel choices are possible. It would also be possible for the host to include additional levels of protocol provided the dependence graph does not have loops. The L3 protocol is CXL.mem where L3 Req maps to M2S-Req, L3-RwD maps to M2S-RwD and L3-Rsp covers the pre-allocated S2M NDR and DRS channels. The dependence graph is a method used to show legal relationships and dependencies in the protocol and between protocols and quickly determine high level deadlock freedom in CXL itself and with host or device internal protocol choices.

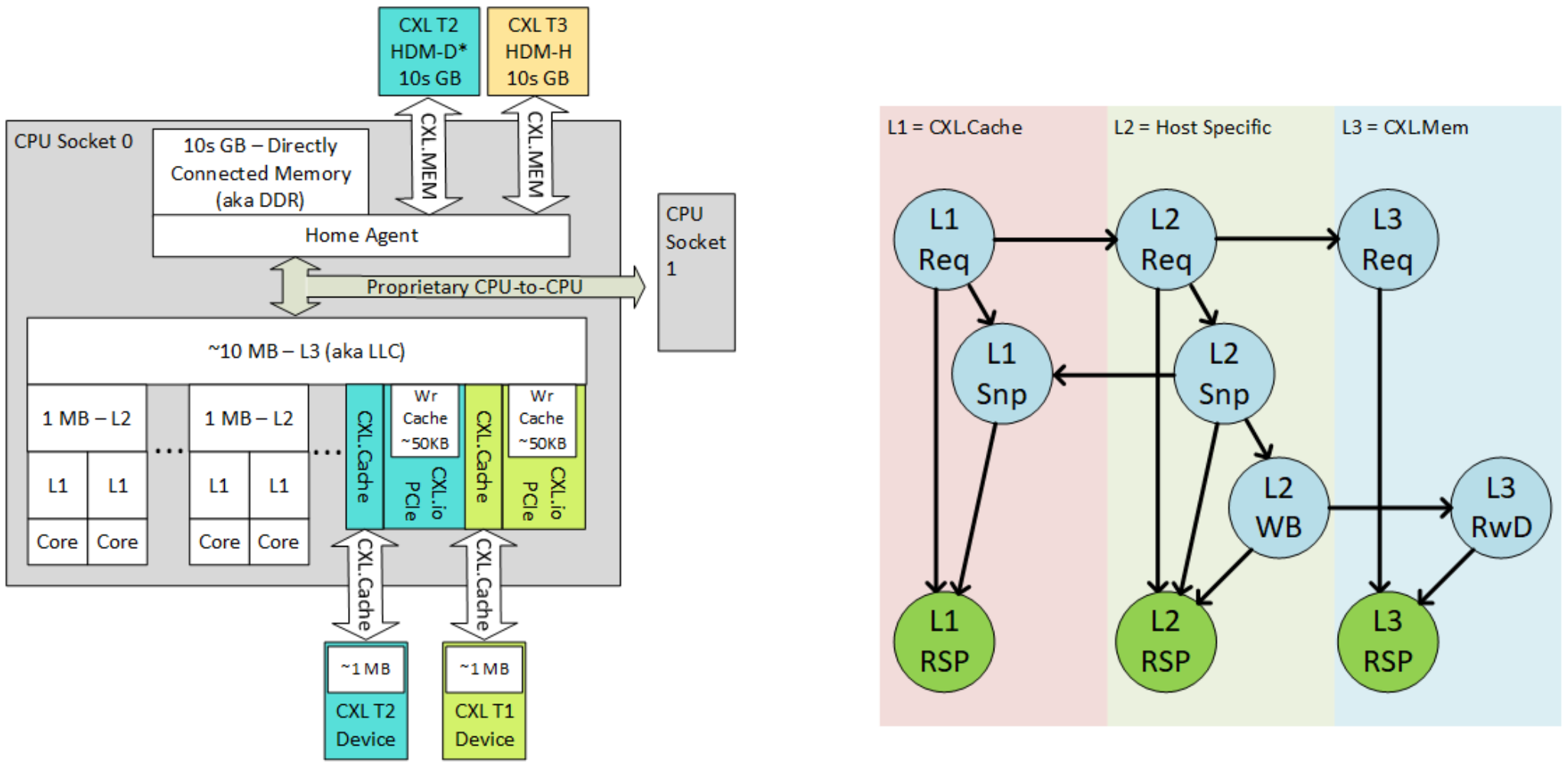

A. Host Cache Construction with CXL.Mem/CXL.Cache

B. Coherence Protocol Dependence Diagram

*Figure 10 CXL Protocol Hierarchy*

## 4 CXL 2.0 PROTOCOL

CXL's second generation enables resource pooling which allows assigning the same resources to different hosts over time. The ability to reassign resources at run time solves resource stranding (Challenge 3 in Section 1) as it overcomes the tight coupling of resources to individual hosts. If one host runs a compute intensive workload and does not use the device memory assigned from the pool, operators can reassign this device memory to another host, which might run a memory intensive workload. As operators typically do not know which workloads run on which hosts at design time, resource pooling can thus save significant memory: instead of sizing the two hosts for the worst-case memory capacity used by any type of memory-intensive workload, operators can provision memory for the average case [18]. The same pooling construct is applicable to other resources like accelerators.

### 4.1 CXL 2.0 Protocol Enhancements

CXL 2.0 adds Hot-Plug, Single Level Switching, Quality-of-Service (QoS) for Memory, Memory Pooling, Device Pooling, and Global Persistent Flush (GPF). Hot-Plug was not allowed in CXL 1.1 which precludes adding CXL resources after platform boot. CXL 2.0 enables standard PCIe hot-plug mechanisms enabling traditional physical hot-plug and dynamic resource pooling.

To support a single level of switching, CXL standardizes address decoding for CXL.mem address regions in HDM decoders. The hierarchical decode follows the PCIe model of memory decode which allows decode at each switch avoiding the need for the host or switch to fully decode all agents in the hierarchy. To support multi-host connections and enable device pooling, each host represents the CXL topology as a virtual hierarchy (VH) which includes the switch and virtual bridges for the host's port and every port with device resources. Each host sees a separate VCS (Virtual CXL Switch) that includes bridges for the devices assigned to this host, as shown in Figure 11. Flits are routed to devices based on the active virtual bridges in the VH. This limits CXL 2.0 to directed tree topologies with at most one path between each host and device. Furthermore, the need to track each VH's address maps in the switch limits scalability to a single switch level CXL 3.0 overcomes this limitation (Section 5.4). Configuring and changing the VH is described in Section 4.2.

Device Pooling builds on top of the multi-host switch support by allowing devices to be dynamically assigned to one host at a time. Standard devices are assigned to a single host at a time and are referred to as Single-Logical-Device (SLD).

CXL also defines a Multi-Logical-Device (MLD) which allows a single CXL.mem device's resources to be divided into logical devices (up to a maximum of 16) that can be assigned to different hosts at the same time. Each logical device within an MLD can be assigned to a different host. CXL 3.0 introduces an extension called Dynamic Capacity Device that is even more flexible. Each logical device (LD) within an MLD is identified by an identifier (LD-ID). This extension is only visible on links between a switch and device and not visible to a host. The host leaves the LD-ID field blank. The CXL switch applies the LD-ID tag to a host's CXL.io and CXL.mem transactions based on the host's port.

As a result of switching, MLDs, and potentially different types of memory (which might not be based on DRAM) resources in a single switch may become oversubscribed resulting in Quality-of-Service problems. For example, low performance in one device may cause congestions for all agents using the switch. To mitigate this problem, CXL 2.0 introduces a DevLoad field into CXL.mem Response messages to inform the host of the load observed in the device which it is accessing. The host is expected to use this load information to reduce the rate at which it sends CXL requests to that device. The CXL specification defines a reference model such that the injection rate is reduced at high or critical load until nominal loading is reached and at light load the host can increase injection rate until nominal load is reached.

If there are multiple CXL request injection points (e.g., in a CXL 2.0 pooled scenario), QoS is provided through source throttling, which controls how much of the MLD resources can be consumed by each source. For non-shared resources, each VH can be isolated by ensuring that transactions for different VHs (or PBR destination) can progress without any inter-dependency. The CXL protocol also defines a containment model which ensures that if an end-point device is not responsive, the impacted VH will contain the error by generating an error response for outstanding accesses within the host to avoid host timeout and could otherwise bring down the VH.

CXL also defines error reporting for memory devices, including poison data support. While the exact error correcting code depends on the type of memory media as well as the usage model and platform, reporting is defined for both corrected as well as detected but not corrected errors. The reporting has been standardized for software to take necessary action [3, 5].

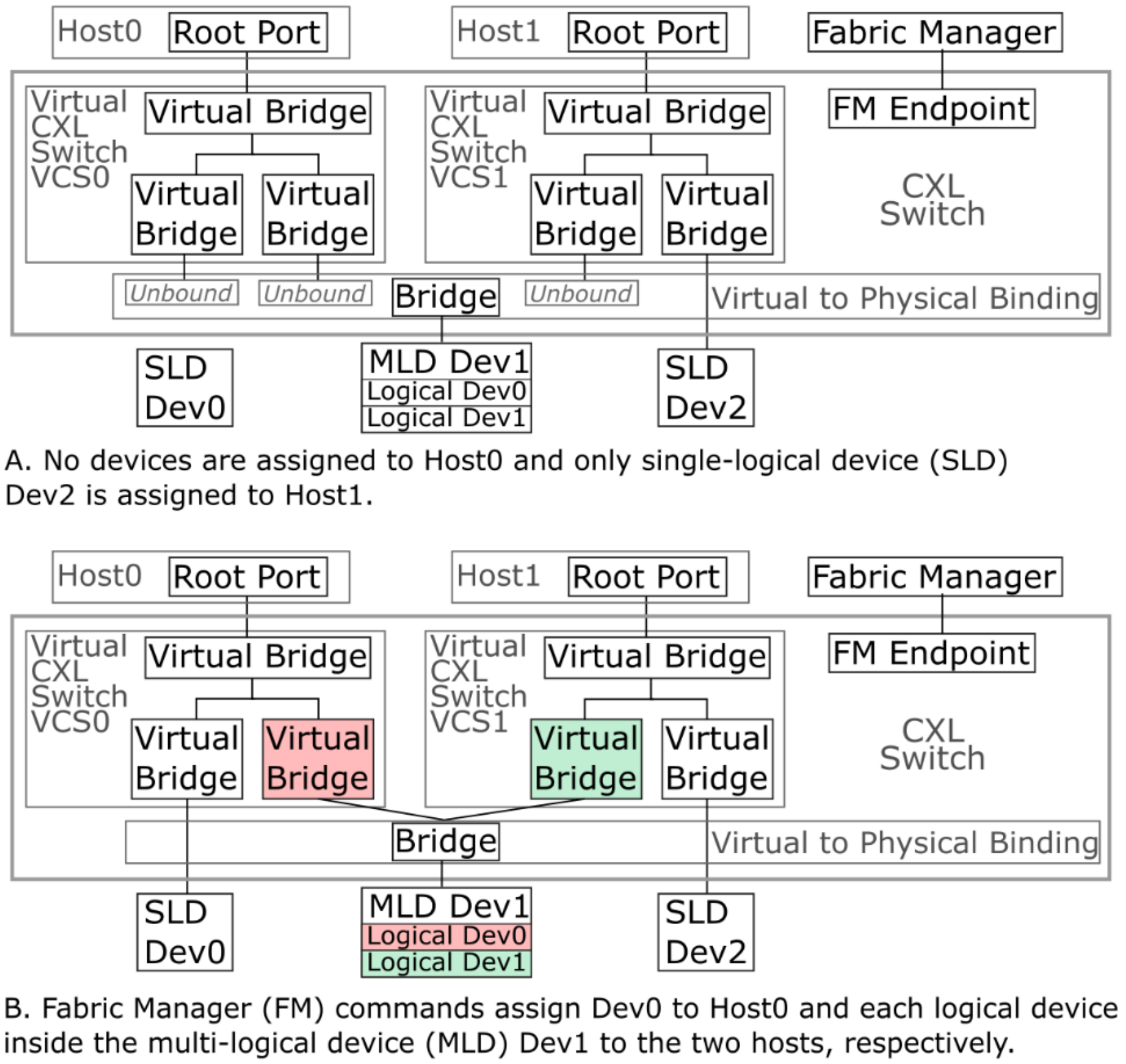

*Figure* 11. *Example of a CXL switch with two virtual hierarchies*

### 4.2 Pool management

Pool management refers to the assignment of CXL devices to hosts at run time. The standard uses the term CXL Fabric Manager (FM) to denote the application logic and policy that makes assignment decisions. In practice, the FM can be software running on a host, firmware embedded within a Baseboard Management Controller (BMC) or within a CXL switch, or a dedicated device. The FM assigns a (logical) device to a host by using the Component Command Interface (CCI). This can happen either in-band, e.g., via Memory Mapped Input/Output (MMIO), or out-of-band via the Management Component Transport Protocol (MCTP). MCTP supports many types of out-of-band communication including SMBus/I2C, PCIe, USB, and serial links.

Device assignment to a host is done by the FM via the bind command which takes four parameters: the CXL switch ID, the virtual bridge ID, the physical port ID, and the logical device ID, which is empty for SLDs. The FM sends this command to the specified switch, which checks if that physical port is currently unbound. If unbound, the switch updates its internal state to execute the bind. It then sends a hot-add indication to the host and finally notifies the FM of successful binding. The virtual bridge and the associated device now appear in the host's virtual hierarchy. This transition is shown between Figure 11A and Figure 11B.

For MLDs, the FM may need to first configure a logical device via the Set-LD command. This command creates LDs based on a memory granularity (e.g., 256MB) and multiple memory ranges. When the FM does not have an out-of-band connection to the MLD, it can tunnel this command via a CXL switch. After creating LDs, the FM can assign them to a host using the bind command with the corresponding LD-ID.

To undo a device assignment, the FM uses the unbind command which takes three parameters: the CXL switch ID, the virtual bridge ID, and an unbind option. The unbind option indicates how to execute the unbinding, including options to wait for the host to deactivate the port, to hot-remove and wait for the host, or force hot-remove. With the last option, the FM is able to work with uncooperative hosts, e.g., due to malfunction or because they are assigned to a third-party customer in a bare-metal cloud.

The FM supports 19 other commands in addition to the three described here. They include QoS controls such as bandwidth allocations among logical devices within an MLD. These commands also include ways to identify switch ports and devices, and to query and configure their state (see Table 205 in [3]).

### 4.3 Host Software Support

Even as much of CXL is implemented in hardware and software external to the host (such as the FM), host software plays a major role in supporting CXL. At a high level, system firmware enumerates and configures CXL resources that are present at boot time. The operating system (OS) enumerates and configures CXL resources that are attached and detached at run time, e.g., memory address ranges that are hot-added and hot-removed. Thus, the OS plays a major role in supporting CXL 2.0.

Traditionally, resource topology and affinity are communicated via static ACPI (Advanced Configuration and Power Interface) tables. For example, NUMA (Non-Uniform Memory Access) domains are defined in the SRAT (Static Resource Affinity Table) table and memory bandwidth and latency are defined in HMAT (Heterogeneous Memory Attribute Table) tables. Typically, these tables are configured by firmware which had knowledge of all resources at boot time[5]. In PCIe, this is a simpler problem as plugging devices does not affect system-mapped memory. In CXL, memory-specific resource characteristics need to be communicated for memory that is attached at run time. With an ecosystem of CXL vendors and devices, it is not feasible to define configurations in advance. CXL 2.0 introduces the Coherent Device Attribute Table (CDAT) which describes internal NUMA domains, memory ranges, bandwidth, latency, and memory usage recommendations [47, 48]. When a memory (logical) device is hot-added at run time, the host reads the associated CDAT registers and the OS assigns a free HPA (Host Physical Address) range and programs the HDM decoder.

## 5 CXL 3.0 PROTOCOL

To meet data sharing challenges in large distributed systems (Challenge 4), CXL 3.0 expands the resource pooling introduced in CXL 2.0 to a much larger scale with multi-level switching and protocol support for up to 4096 end devices, taking advantage of the larger Flit size, with low-latency and sufficient bandwidth. An end device can be a host CPU, representing an independent server or node, memory, accelerator, or any other I/O device such as a NIC. The overarching goal is dynamically composable systems depending on the workload to deliver power-efficient performance with lower TCO, as shown in Figure 12. CXL 3.0 introduces the following to meet these goals:

- Doubling the per-pin bandwidth while keeping latency flat required for scaling to larger topologies.
- Support for Fabric topology: This is a first for any load-store interconnect standard which has ordering constraints among transactions. With multiple paths between any source-destination pair, CXL breaks away from restrictions of the tree-topology, which is essential to scale to thousands of devices. This enables lower latency, higher bisection bandwidth, and fail-over capability.

---

[5] In CXL 1.1, firmware can enumerate and configure devices as in traditional PCIe. Specifically, firmware uses the CXL Early Discovery Table to retrieve pointers to host bridge registers, memory windows, to set up memory address maps, and program HDM Decoders [47].

- Direct peer-to-peer access from a PCIe/CXL device to the coherent HDM memory hosted by a Type-2/ Type-3 device without involving the host processor if no conflict arises. This results in low latency, less congestion, and high bandwidth efficiency which is critical for large systems. For example, in Figure 12, the NIC to memory is 8 hops away using direct P2P vs. going through a CPU is 16 hops away round-trip.
- Shared coherent memory and message passing among across hosts: This can be enforced by hardware or software. Shared coherent memory enables multiple systems to share data structures, perform synchronization, or pass messages using low-latency load-store semantics. Message passing can also be done by using the load-store CXL.io semantics (vs. networking semantics with higher latency).
- Near-memory processing to allow computation to be performed near memory for better performance and energy efficiency.

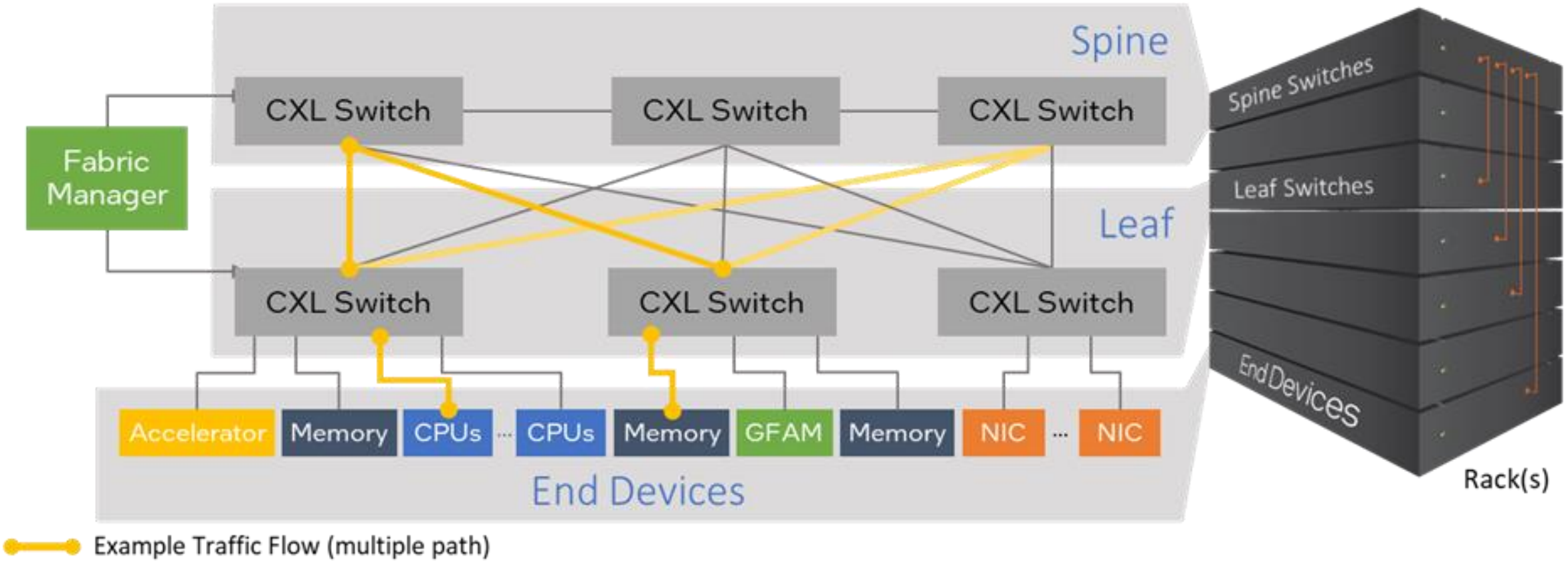

*Figure 12: Fabric topology with multiple-paths enabled by CXL 3.0 spanning one or many racks, enabling composable scale-out systems with shared memory and load-store message passing among peers*

### 5.1 5.1 Flit Formatting to support doubling of Data Rate to 64.0 GT/s

CXL 3.0 supports 64.0 GT/s using PAM-4 (pulse amplitude modulation 4-level) signaling, based on the PCIe 6.0 specification [5,6,7]. With 4 voltage levels to encode two bits per unit interval (UI), PAM-4 results in very high error rates due to reduced eye height and width (i.e., voltage and time) [13, 14]. To overcome this, PCIe 6.0 specification adopted a first burst error rate of 10-6 and introduced a 256-byte Flit (Figure 13a), where the payload comprising of 236B of TLPs and 6B of data link packets (DLPs) are protected by an 8B CRC based on the Reed-Solomon code over GF($2^8$) and the entire 250B is protected by a 3-way interleaved single-symbol correct forward-error correcting code (FEC) of 2B in each way [7, 14]. CXL 3 at 64.0 GT/s uses the same electrical layer with a 256-B Flit, but with some modifications. It retains the same FEC mechanism as PCIe for the two variants of 256B sized Flits, as shown in Figure 9b and Figure 9c.

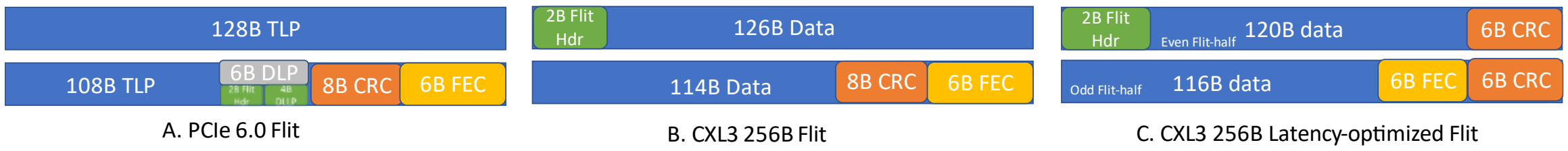

*Figure 13: Three types of CXL 3.0 Flits. A 256B latency-optimized Flit (C) consists of two Sub-Flits of 128B each with common FEC across 256B but separate CRCs.*

Figure 13b shows the layout of the regular 256B Flit in CXL 3. The 8B CRC is identical to that of PCIe 6.0. The only difference is the 2B Hdr (Header, which is an enhanced version of the first two bytes of DLP, which is used for reliable Flit delivery management such as acknowledge, replay request, replay with sequence numbers) is enhanced to indicate the type of Flit (CXL.io, CXL.cache+mem, ALMP, Idle) along with the reliable Flit delivery management control and is placed in the start of the Flit to pipeline the delivery of the Flit to appropriate protocol stack without accumulating the entire 256B Flit to reduce latency. An optional latency-optimized (LO) 256B Flit, as shown in Figure 13c. The LO Flit is subdivided into two sub-Flits, each 128B, with the FEC in the odd Flit-half and the 2B Flit Hdr in the even Flit-Half. A 6B CRC is present in each Flit half. This 6B CRC is derived from the 8B CRC for reduced gate count, as described in [5, 13, 14]. The first 6B CRC protects the 2B Flit Hdr and 120B data in the even Flit-half and the second 6B CRC protects the 116B of data in the odd Flit-half. An even Flit-half can be processed in order if its CRC passes without waiting for the odd Flit half. The odd Flit-half can be processed if its CRC passes, assuming the even Flit-half has been processed (i.e., no CRC error). Since the CRC is about 10-levels of logic gates vs. 50 levels of logic gates for FEC [14], applying CRC first helps reduce the latency by 2 nanoseconds on a x16 link [13]. Also, the CRC over 128B makes the accumulate latency slightly better than the 68-B Flit accumulation latency at 32 GT/s. If an error is detected, then the entire Flit is accumulated, the FEC applied, and then the CRC is applied again. Details of the Flits and the FEC and CRC mechanisms can be found in [5, 7, 13, 14].

The type of Flit to be used (68B, 256B, LO) is negotiated upfront when the CXL protocol is negotiated with 8b/10b encoding. 68B Flit support is mandatory for any CXL device. If the CXL device supports 64.0 GT/s data rate, it must also advertise the 256B Flit mode while the advertisement for the LO Flit mode support is optional. If all components of a link (the two devices as well as any Retimers in between) support the LO Flit mode, then LO is selected; else if all components support 256B Flit, then the 256B Flit mode is selected; else 68B Flit mode will be supported. Once a Flit mode is selected during early negotiation, that is used irrespective of the data rate of operation. Suppose we selected the LO Flit mode because all components in the link support it along with 64.0 GT/s but if the link is operating at 32.0 GT/s (e.g., power savings speed down-shift or link stability issues at 64.0 GT/s), the LO Flit mode will still be used.

For CXL.io, the last 4B of the 'Data' will be for DLLP (the equivalent of the last 4B of the 6B DLP in PCIe 6.0 [7,14]. That leaves 236B for TLP in 256B Flit mode (same as PCIe 6.0 Flit) and 232B for TLP in the LO Flit mode. For ALMP, most of the 'Data' field is Reserved. For CXL.cache-mem, the slots arrangements are shown in Figure 14. The H- and G-slots have the same usage as 68-B (i.e., H-Slot or Header Slot is used for Headers only whereas G-Slot or Generic Slot is used for either Header or Data) with the exception that the additional 2B in H-slot is used for building larger topologies. The HS slot is 10B only and used for small headers like 2DRS or 2NDR.

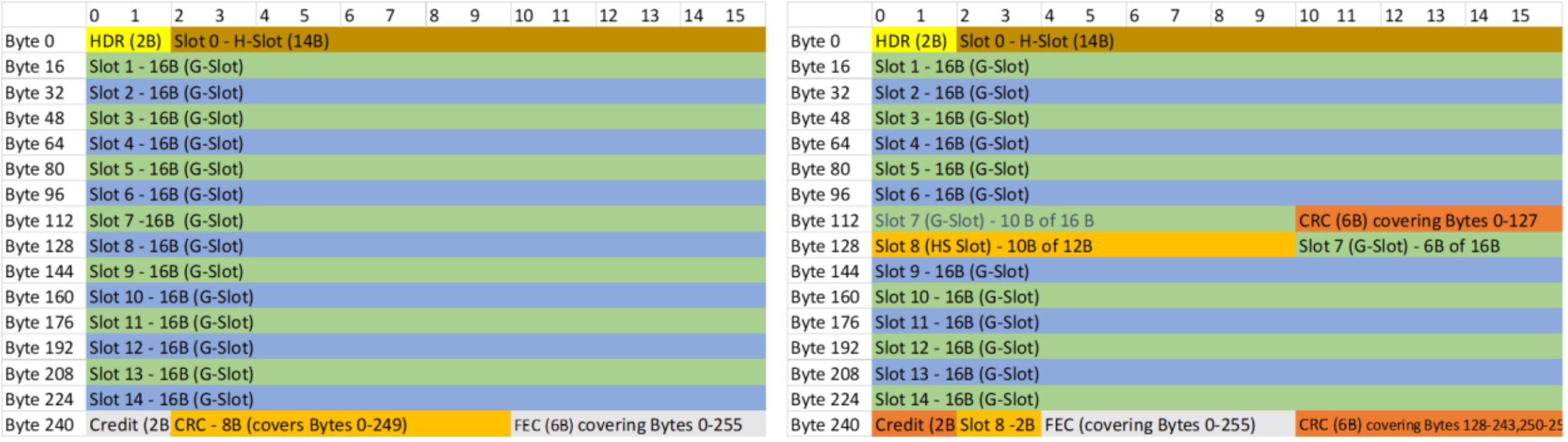

(a) 256-Byte Standard Flit Layout

(b) 128-Byte Latency Optimized Flit Layout

*Figure 14: CXL.cache and CXL.mem slots for 256B and LO Flits [8]*

### 5.2 Protocol extensions with CXL 3.0: Unordered I/O (UIO) and Back-Invalidate (BI) for fabric support

The PCIe Ordering rules described in Section 3 preclude any non-tree topology, which means that only one path exists between any two nodes (host or device) in the CXL network. We need redundant and multiple paths between any source-destination pair of nodes to create large distributed systems with good performance. Since software relies on the producer-consumer ordering model, the CXL protocol needs to preserve that while accommodating non-tree topologies. CXL 3.0 (and hence PCIe) solves this challenge by introducing 'unordered' read/ write/ completion transactions on one or more distinct virtual channels (VCs, see Section 3.1) and transferring the ordering enforcement to the source node only. Each VC is independent of other VCs and each VC comprises 3 three flow control (FC) classes: Posted (P) for transactions such as Memory Write, Non-Posted (NP) for transactions such as Memory Read, and Completions (C) for completions to each NP transaction, as described in Section 3.3. The basic idea is that every UIO Write gets a completion (`UIO Write Completion' in the C FC class). Hence, UIO Writes are fundamentally non-posted even though it is on the P FC class to enable the source to enforce ordering. Thus, the producer (e.g., Device X in Figure 6a) must wait for the completion of all the `Data' before writing to the `Flag' signaling that the data is available. An UIO Read is like a regular memory read in that it gets one or more completions (`UIO Read Completion' in C FC Class) with or without data along with the status, including errors. Thus, in UIO VC, P and NP have only one type of transaction each (UIO Write and UIO Read respectively) and C has 3 types of transactions (UIO Write completion – no data, UIO Read Completion with data, and UIO Read Completion – no data). Furthermore, transactions across the FCs as well as within an FC have no ordering requirements, as summarized in Figure 15. This enables CXL.io (and PCIe) transactions to be sent on any path in a topology with multiple paths between a source-destination pair and still enforce the producer-consumer ordering semantics. VC0 will always be used for the traditional non-UIO Ordering for backward compatibility following the tree topology for routing and one or more VCs in VC1-VC7 may be used for UIO traffic that can use any of the paths for any transaction.

| 1st Trans / 2nd Trans | Posted (P) (UIO Write) | Non-Posted (NP) (UIO Read) | Completion (C) (UIO Rd/Wr Cpl) |
|---|---|---|---|
| P (UIO Write) | Y/N | Y/N | Y/N |
| NP (UIO Read) | Y/N | Y/N | Y/N |
| C (UIO Rd/Wr Cpl) | Yes* | Yes* | Y/N |

(*: Completions must not be blocked by P/NP to avoid deadlock. Transactions flow completely unordered within and across different FCs in the UIO VC)

*Figure 15: UIO Ordering Rules. A 'Y/N' indicates that the 2nd transaction may or may not bypass the 1st transaction.*

CXL 3.0 introduces a new flow with two channels in CXL.mem: Back-Invalidate (BI) in the S2M direction and the response BI-Rsp in the M2S direction, as described in detail in Section 5.2. This gives rise to a new type of device memory: HDM-DB (Host-managed Device Memory – Device-coherent with Back-invalidate support), supported by Type-2 and Type-3 devices.

### 5.3 Peer-to-peer communication between devices and mapping large chunks of local memory

The Back-Invalidate flow in CXL 3.0 enables three sets of usages: direct peer-to-peer communication between a CXL/PCIe device through CXL.io UIO access to HDM-DB memory (as shown in Figure 16), the ability of Type-2 devices to implement a snoop-filter and map a large chunk of local memory to the HDM-DB region (as shown in Figure 17a), and hardware-enforced coherent shared memory across multiple independent hosts (Figure 17b). The rationale for having BI in CXL.mem is to ensure that there is no other dependency which is required for deadlock avoidance.

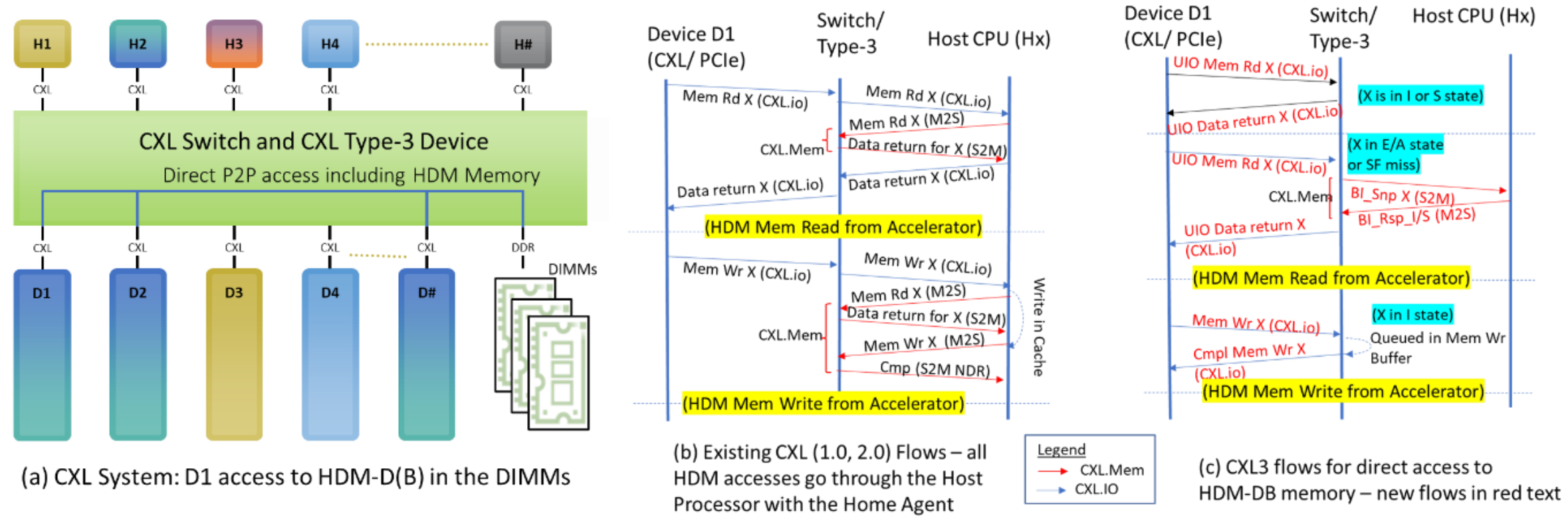

*Figure 16: CXL 3.0 Protocol Enhancements with UIO and BI for latency and bandwidth optimization [9]*

With UIO and BI, a device can directly access HDM-DB memory without going through the host CPU. An accelerator device connected to a switch that needs to access memory connected directly to the switch (Figure 16a) that is mapped to the HDM-D or HDM-H region needs to go through the host CPU (Figure 16b) with CXL 2.0 flows which involves the host getting the data from memory and resolving coherency flows prior to completing the access from the device. However, with CXL 3.0 flows, the device sends the same memory read and write transactions using CXL.io in the UIO VC. The switch does not route these UIO transactions to the host. UIO transactions are serviced directly from memory, if the state of the cache line is I or S for a UIO Read or I for a UIO Write; else the memory controller back-snoops the host processor using the BI flows, as shown in Figure 16c. As we will see later, even when BI is needed, this new mechanism is more bandwidth efficient than the existing approach of sending all requests to the host which caused more traffic and extra latency.

UIO combined with BI enables an I/O coherent model which can scale to lots of devices. Those devices that need to cache memory can still use CXL.cache semantics when they need it and use UIO for the rest. This reduces the snooping overhead as well as the latency involved in going to the host processor all the time. The second aspect of BI is to enable Type-2 devices to have a snoop filter instead of having a full directory and invoking the existing bias-flows. This is demonstrated in Figure 17a where the Type-2 device invokes the BI flows to evict a cache line from its snoop filter if a new request encounters a capacity miss.

The BI-mechanism also enables the implementation of shared and hardware-enforced coherent memory across multiple hosts, as illustrated in Figure 17b. Here the memory device (GFD/MLD) maintains a directory (or a snoop filter) to track ownership of cache lines in the coherent shared memory. Thus, when Host 1 obtains ownership of cache line X as a shared copy, it updates the directory from ‘I’ state to ‘S’ state with Host 1 as the sharer. When Host 3 asks for the same cache line X in shared state, it provides the data and updates the directory for X to indicate that both Host 1 and Host 3 have it Shared. When Host 4 requests an exclusive copy, it issues a Back Invalidate to both Host H1 and Host H3, waits for the response from both to ensure that Host 1 and Host 3 have invalidated their copies of X and updates its directory to mark X as ‘E’ by Host 4, prior to sending the data and ownership to Host 4. This memory device can be a Multi-Logical Device (MLD) or a GFAM (Global Fabric Attached Memory) Device (GFD). Each GFD can scale to support up to 4096 independent nodes (vs. 32 for MLD) simultaneously for either pooled or shared memory by not being discoverable by each node independently through configuration space.

To support shared coherent memory across multiple independent hosts, devices can implement an on-die snoop filter and/or a directory structure in memory. Devices can use two bits for coherency states (Invalid, Shared, Exclusive) followed by a sharing list of hosts. The implementation for this list is up to the device. As an example, for invalid, the list entry is ‘don’t care’. For exclusive, the list contains the host ID that has it exclusive. For sharing, one can implement a combination of bit-vector, with each bit representing one or a group of hosts or a list, where the list can identify each hosts up to a certain number beyond which it can be a coarse grain representation of host groups. Implementation details will depend on how many directory bits are available, and the number of hosts that can access a cache line. One may also choose to maintain this snoop-filter / directory across multiple cache lines (e.g., at a page level). The size of shared memory and the number of hosts that can share a cache line depends on the usage model. Examples include large in-memory applications such as very large databases, logs, machine learning, key-value store, in-memory analytics, etc. Hardware coherence can scale for these applications where it is mostly read-only. In some cases, software coherence may be desirable. The CXL3 architecture also enables inter-domain interrupts, semaphores using shared memory, data replication, and message passing using special address map regions through shared memory controllers, described in detail in [9].

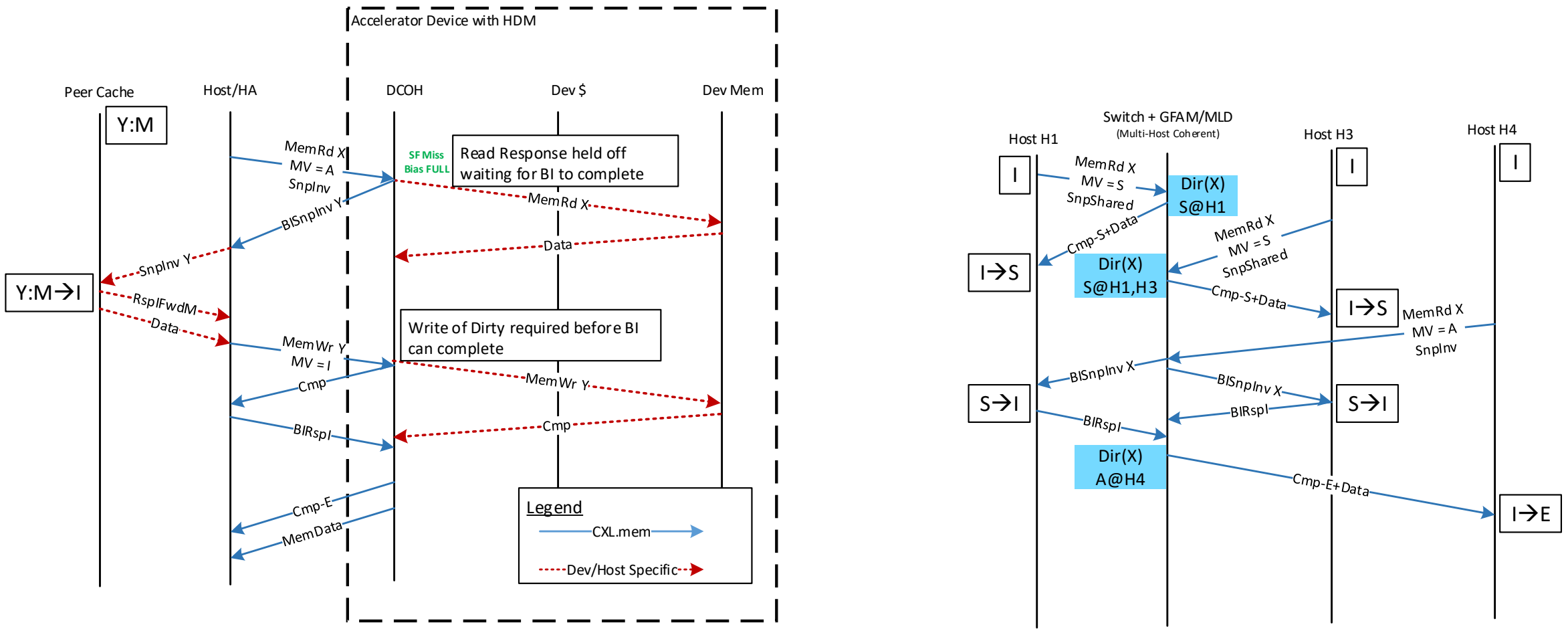

A: Snoop-Filter Capacity Miss Flow with BI

B: Shared Memory coherency across Multiple Domains with BI

*Figure 17: (a) Existing Bias – Flip mechanism needed HDM to be tracked fully since device could not back snoop the host. Back Invalidate with CXL 3.0 enables snoop filter implementation resulting in large memory that can be mapped to HDM, (b) The Back Invalidate Flows of CXL 3.0 can be used to implement hardware-based coherent shared memory across multiple hosts, each with its independent coherency domain*

Back-Invalidation creates two additional channels in the protocol and link layers: BISnp and BIRsp. These channels are necessary to ensure the protocol deadlock freedom in the updated dependence diagram show in Figure 18 adding BISnp (S2M BISnp) compared with the CXL 1.1/CXL 2.0 Dependence graph in Figure 10-B. The BIRsp is pre-allocated so for the dependence graph it is combined into L3 RSP in the graph.

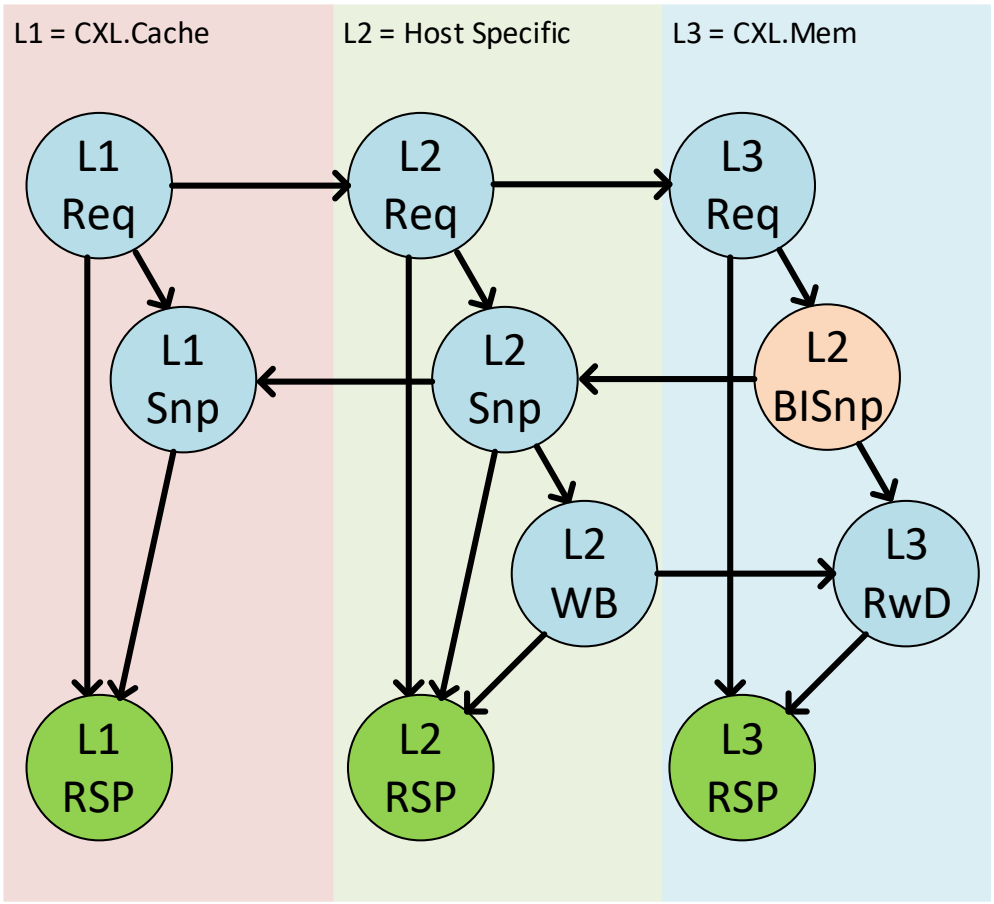

*Figure 18 Updated CXL 3.0 Protocol Dependence with Back-Invalidate*

### 5.4 CXL.cache Device Scaling

Prior to CXL 3.0, each Virtual hierarchy could only support a single CXL.cache device. To enable additional devices an additional 4-bit ID space 'CacheID' was added to the messages to enable up to 16 CXL.cache devices to each root port. For the host to support CXL.cache it must have snoop filter structure that will track each address that the device may be caching. The size of this tracking in the host will limit the device's ability to cache host memory, so sizing in the host will constrain how much host data the device may cache. Note that the host does not limit the device's ability to cache its own memory for Type-2 devices. With the addition of CacheID, the host must individually track each device behind a root port to avoid degrading to a multi-cast snooping behavior. Multi-cast snooping would be functional but would not meet the bandwidth/performance expectations that devices should rarely see snoops for an address that are not in the device cache. The constraints of coherence tracking in the host may limit the number of CXL.cache devices in total that can be supported. The host will advertise this limit at the host bridge granularity. Software can discover this capability and will enforce the limit during the enumeration of CXL hierarchy.

### 5.5 Port Base Routing (PBR) and Extensions for CXL Fabric

To enable scaling to 4096 endpoints (hosts or devices) and non-tree (multi path) network topologies, CXL 3.0 adds a new optional protocol format called Port Based Routing (PBR). The idea is that PBR is a simpler and more scalable protocol than CXL 2.0's hierarchical routing (Section 4.1) by focusing only on information that's needed to route messages between PBR switches. For example, hierarchical routing would require the switch to know the virtual hierarchy and the address mappings for every host. CXL 3.0 continues using standard CXL 2.0 messages for the link between most endpoints and the first switch (e.g., Leaf to endpoint links in Figure 12). This first switch port connecting to device or host is called 'edge port' because it has a special role in converting between standard messages (HBR) and PBR messages. PBR routing itself is used for inter-switch links, e.g., Spline to Spline and Leaf to Spline links in Figure 12). From a software point of view, only the fabric manager needs to configure the network of switches; software running in the host just sees a flat topology that ends at the first switch Port.

PBR addresses endpoints with a 12-bit ID referred to as PBR-ID (PID). The edge port adds this identifier to messages as a Destination PID (DPID) and in some cases a Source PID (SPID). The generation of the PID is done by translating/decoding the HBR message to/from PBR message using various methods. This translation is stateless, i.e., does not require tracking any outstanding requests. Examples of this translation include Address to PID, LD-ID to/from PID, CacheID to/from PID, Bus Number to/from PID. For LD-ID to PID, the PID offers fewer bits which is solved by a 16-deep lookup table that is defined with the 4-bit LD-ID indexing the table to determine the PID of the host. In the case of Address to PID this is done by decoding the Host Physical Address (HPA) with a lookup of the Fabric Address Segment

Table (FAST). To improve scalability the FAST provides a single memory range which gives a power of 2 size to each PID which enables direct use of selectable address bits to lookup the PID owning the region of HPA. As the HBR message to/from PBR message is transparently performed by the edge port, hosts and devices can work with PBR without being directly aware of PBR. Furthermore, since address decoding has already been performed at the edge, interior PBR switches (e.g., Spline switches in Figure 12) can route messages purely based on the DPID and do not have to perform address decoding. Without this simplification intermediate links that are shared by multiple virtual hierarchies would need to have the full address decode capability for all Virtual Hierarchies (VHs) that share the link and thus would limit the number of VHs. PBR thus enables better scalability and reduces latency and cost of interior switches.

Routing of PBR messages is centrally controlled by the Fabric Manager (FM, Section 4.2). The FM configures each PBR switch with a lookup table indexed by the 12-bit DPID. These tables maps DPIDs to the switch's outgoing physical port. To enable multi-path routing the lookup table may provide multiple destination physical ports where unordered traffic may be routed dynamically to target ports based on load. For CXL.io the UIO VC is completely unordered and can be identified easily by the VC in which the traffic arrives. CXL.cache-mem is primarily unordered with a few limited ordering exception rules for resolving conflicts (Example: In CXL.Cache Snoops push GO to the same address); these exceptions can be easily enforced even with multiple paths by ensuring the ordered messages follow the same path interleaving appropriately. For example, tag bits in the CXL.mem Non-Data Response channel can be used to interleave as only messages with the same tag require ordering. Ordered traffic such as traditional CXL.io has less flexibility in that the same single path must always be selected which conforms to the tree-based topology across all the links. When the FM observes a link or switch failure, it can also reconfigure the routing table and redistribute new lookup tables to PBR switches.

Inter-switch links (ISLs) carry the PBR message format. These links are also symmetric in nature in that they may carry upstream and downstream traffic from different hosts in opposing directions on the same link. Specifically, a non-PBR port supports six CXL channels up and six CXL channels downstream. A PBR port must be able to carry 12 CXL channels upstream and downstream, respectively. Support for ISLs requires the underlying Cache and Mem link layer to supply a fully symmetric set of channels, each with its own set of flow control credits. This contrasts with device and host links which are either upstream or downstream links (host to switch is downstream and device to switch is upstream),

Instead of relying on edge ports, a memory device may also directly participate in PBR. Such an endpoint is referred to as Global Fabric-Attached-Memory Device (GFD). The GFD allows for high-scalability memory devices that can be shared/pooled across all 4095 other agents in the CXL fabric where CXL 2.0 Multi-Logical Devices (MLD) can only be shared by a maximum of 16 hosts.

## 6 SURVEY OF CXL IMPLEMENTATIONS

The CXL ecosystem requires CPU and device support. While CXL 3.0 products are expected to arrive in a few years, CXL 1.1 and CXL 2.0 are deployed in several commercial products. Intel supports CXL starting on Sapphire Rapids (SPR) CPUs as well as on Agilex7 FPGAs, supporting all three protocols. AMD supports CXL in Genoa and Bergamo CPUs and has announced support for SmartNIC devices [52], ARM has announced CXL 2.0 support in the V2, N2, and E2 series CPUs [57].

On the device side, IP vendors such as Synopsys, Cadence, PLDA/Rambus, Mobiveil, and others have demonstrated interoperability with SPR CPUs [8, 21, 22, 23, 24]. Samsung has built a CXL 1.1 memory expansion device and publicly shared its benchmarking results [54]. Montage [55], SK Hynix [56], Microchip [79], Micron [77], Astera [78] have announced CXL Type-3 memory devices. Micron has prototyped a CXL 1.1 near-memory computing device [53]. Academic studies have also independently reproduced results on SPR [50] as well as implemented custom CXL hardware and software prototypes [51].

In the next sections, we report measurements of CXL implementations. We've tested CXL.mem on Intel and AMD CPUs with multiple device vendors. However, since public performance numbers are primarily available for Intel CPUs, our survey primarily focuses on the performance of the Intel implementation as a representative performance indicator of CXL implementations.

Figure 19 is a representative micro-architecture of CXL IP, with industry standard interfaces such as PIPE [25], LPIF [26], CPI [27], and SFI [28], for CXL endpoints as well as a host-processor such as SPR. The memory size, caching hierarchy, and cache sizes in Figure 10a are meant to provide a scale and not the exact magnitude. The last level cache (LLC) covers the lower levels of cache in the hierarchy as well as the write caches associated with the PCIe/ CXL.io stack inside the CPU. The LLC also covers caches across all CXL devices. A snoop filter provides the home agent with information on when to snoop a given CXL device. Memory, whether locally connected to the CPU (e.g., DDR), or belonging to the CXL device mapped to the system address space, is under the purview of the Home Agent (represented by blue and green colors in Figure 10a).

### 6.1 Measured Latency of CXL 1.1 implementations

At a high level, CXL latency is composed of a protocol component and a queueing component, which depends on load. We first focus on CXL protocol latency, in an idle system. Figure 19 shows the latency breakdown on Intel SPR for different CXL blocks. The PHY block shows the analog circuitry responsible for serializing / deserializing the data. It connects to the PHY Logical block through the parallel interface of PIPE (32 bits at 1GHz per Lane for a 32.0 GT/s link). The PHY Logical block also performs link training, equalization, (de)scrambling, precoding, lane reversal, link width degradation, polarity inversion, clock domain crossing, clock compensation, de-skew, and physical layer framing.

The LPIF die-to-die interface [26] connects the PHY Logical to the Arb/Mux block and from there to the multiple link layers. The Arb/Mux block performs the arbitration and multiplexing between the two stacks (CXL.io and CXL.cache+mem, see Section 3.2) and also virtualizes the physical layer to the independent link layer stacks for power management [1,2,3,8].

The Link layer (LL in Figure 19) performs CRC checks, manages credits, and handles link level retry whereas the transaction layer (TL) performs the Flit packing/ unpacking, stores transactions in the appropriate queues, and processes the transactions. In CXL 3.0 implementations, when 256-Byte or 128-Byte Latency-optimized Flit is negotiated, the Logical PHY will perform the CRC and replay functionality which is common across all stacks and the Link Layer with each stack will continue to provide the same functionality only for the 68-Byte Flit mode for that stack for backwards compatibility. The CXL.io/PCIe transaction layer queues the transactions, processes them, and enforces producer-consumer ordering [8].

Figure 19 applies to CXL.cache+mem path and includes both transmit and receive paths. Multiple optimizations improve latency over a standard PCIe physical layer logic implementation. These include bypassing the 128b/130b encoding (if negotiated during link training [3]), bypassing the logic and serializing flops needed to support degraded mode (if one or more Lanes are bad or if invoked for power management), bypassing the deskew buffer if the lane to lane skew is less than half the internal PHY Logical clock period, adopting a predictive policy of processing entries from the elastic buffer (vs. waiting for the clock domain synchronization handshake for every entry), etc.. The PHY and PHY Logical latency differ depending on whether a common reference clock or independent reference clocks are deployed. If the components are within a single chassis, we expect a common reference clock with a 15 nanoseconds latency. However, if the link connects across chassis, we expect independent reference clocks, with a 4 nanoseconds latency penalty due to clock crossing as well as sync Hdr overheads.

The total latency from the SERDES pin to the internal application layer (CPI or equivalent interface) and back is 21 ns in common reference clock mode and 25 ns in independent reference clock mode. One IP vendor has also reported a round-trip latency of 25 ns [15]. Minor latency variations are expected across implementations due to factors such as propagation delays due to placement of the various blocks, process technology, and the type of PHY design. For example, an existing ADC (Analog to Digital Convertor)-based PHY would likely have an additional latency of about 5 ns over a DFE/CTLE (Decision Feedback Equalizer/ Continuous Time Linear Equalizer)-based Receiver design.

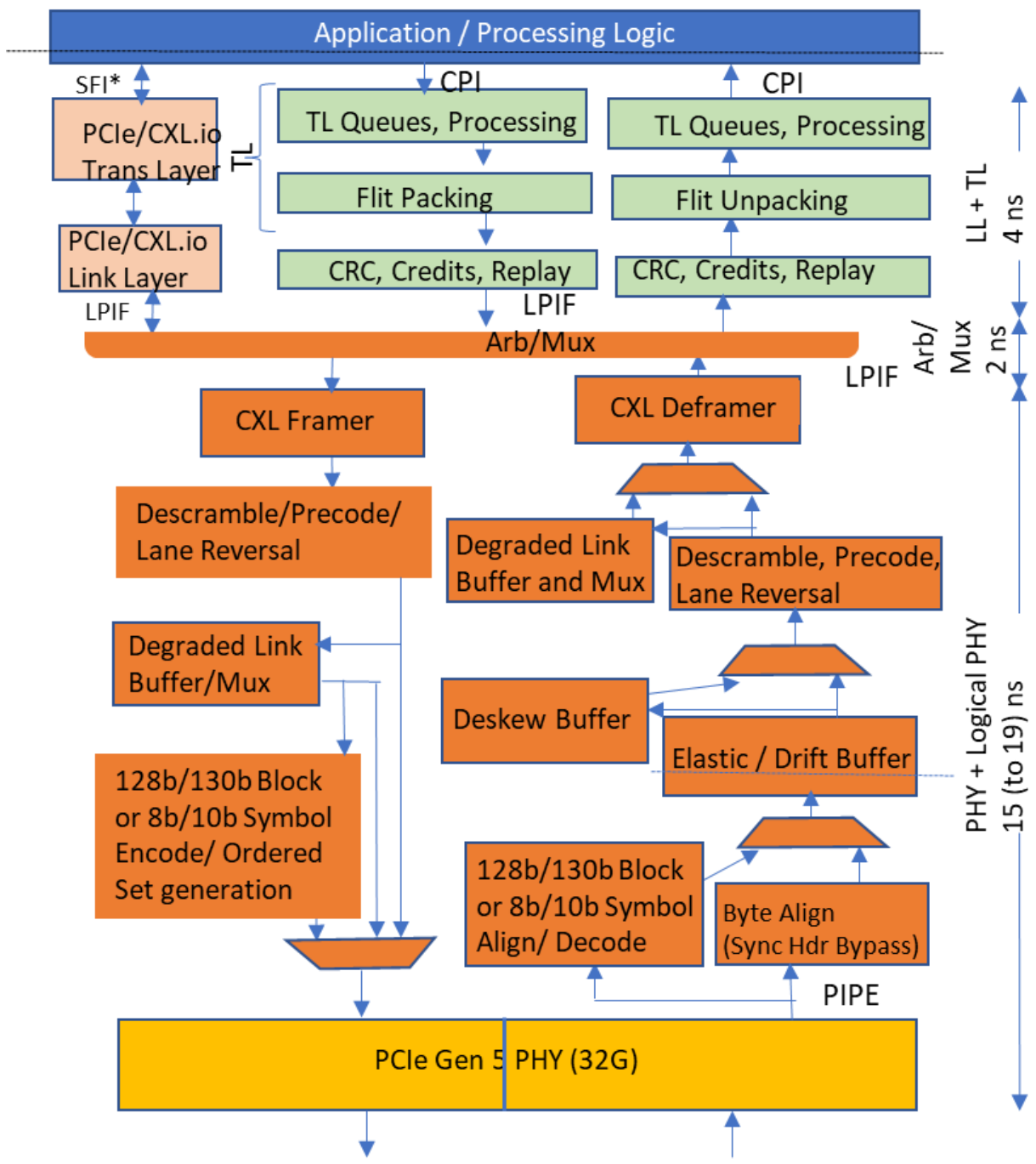

*Figure 19: Representative micro-architecture with typical latency of CXL micro-architecture [8]*

The CXL port's 21-25 nanoseconds round-trip latency occurs twice on a CXL.cache+mem device access, once on the CPU and once on the device. Additionally, we estimate a conservative 15 nanoseconds round-trip flight time with re-timers [17,18]. This gives an end-to-end latency adder of 57 nanoseconds for a memory access across a CXL link. Even with design variations, the CXL specification pin-to-pin latency targets of 80 nanoseconds for a CXL.mem access targeting DRAM/ HBM memory or a 50 nanoseconds response for a snoop [3,4,6] are thus achievable in practice. This adder is comparable to memory access across a CPU-CPU cache-coherent link and well within the latency operating points of CPU [20]. For CXL.io, as in PCIe, the pin-to-pin round-trip latency for a memory read to local memory to completion in Xeon is about 275 ns with an LLC miss and IOTLB hit (with a device having its TLB mandated by CXL, we expect the latency to be like an IOTLB hit).

### 6.2 Projected Latency of large system topologies with CXL 3.0 and beyond

Larger topologies with CXL are possible with switches as well as devices such as `Shared Memory Controller' (SMC), which is a combination of CXL switch along with several memory controllers [9] (Figure 16a), with multi-domain capabilities and PBR routing for scale-out. A CXL switch latency adder to round-trip is expected to be 2x21-25+10ns for internal arb/look-up+10ns flight time on wire = 62-70ns [18, 58]. The various latencies for CXL.cache and CXL.mem for various access/ system configurations are summarized in Table 2.

*Table 2: Estimated latency for CXL.cache and CXL.mem for various topologies [10]*

| Access Types | Estimated Latency |
| --- | --- |
| 1. CPU to Type-3 device (SLD) connecting DDR memory | 170ns |
| | CPU side load-to-use: < 100ns [16] (which includes DRAM access) + CXL |
| 2. CPU to pooled / shared memory on a direct-attach multi-headed (MLD) Type-3 device for DDR memory or SLD connected through an SMC<br><br>3. Device to direct connect memory on host processor across a CXL link w/ CXL.cache (assuming similar load-to-use pipeline as a CPU on device) | stack round-trip of 50ns + 10ns flight time + 10ns Retimer round-trip [17, 18, 58], if used |
| 1. CPU to Type-3 memory through a CXL switch (SLD or pooled/ shared MLD memory / GFAM memory)<br><br>2. Device caching latency to a Type-3 memory connected through CPU expected to be similar for load to use | 250ns<br><br>CPU side load-to-use: <100 ns + CXL switch w/ link flight times 70ns + CXL Type-3 device < 80ns [1, 2, 18, 58] |
| Message to a peer CPU / Device on local SMC or through a CXL switch<br><br>(Assumes UIO based CXL.io for write – completion) | 220ns<br><br>170ns for two CPUs, similar to a local access + 40ns for the switch/ SMC + 10ns for 1-way flight w/ retimer |
| Message to a peer CPU / Device on remote SMC or through 2 CXL switches<br><br>(Assumes UIO based CXL.io for write – completion) | 270ns |

### 6.3 Realizable Data Bandwidth

A x16 CXL link has a *raw* bandwidth of 64 (128) GB/s per direction at 32.0 (64.0) GT/s data rate. However, there are protocol overheads due to the link transferring data besides the actual payload including headers, CRC, and protocol ID/ Flit HDR (see Section 3.2). Additionally, different workloads lead to different request and response patterns and use up Flit slots differently. Thus, the *realizable* data bandwidth is determined by a multi-layer interaction between workload, protocol, and underlying IP. We first discuss overheads that are common across the three protocols (CXL.io, CXL.cache and CXL.mem) and then the *realizable* data bandwidth with common traffic mixes.

For CXL.io in 68-Byte Flit mode, assuming a 2% DLLP overhead, link efficiency is 0.906 and 0.92 with sync Hdr on or off, respectively [8]. Note that native PCIe has a 6% higher efficiency since PCIe does not contain the Flit overheads. For CXL.io with the 256B and 128B LO Flits, the Flit overhead is 6B (Hdr, DLLP) + 6B (FEC) + 8B for 256B / 12B for 128B LO for CRC [10,11]. This is a total of 20B for 256B Flit and 24B for 128B LO Flit with link efficiency of 0.92 and 0.91, respectively[8, 9].

CXL.cache+mem in 68-Byte Flit mode (CXL 1.1-2.0) has three common overheads: 128/130 represents the sync Hdr overhead (unless sync Hdr bypass is supported), 374/375 represents the bandwidth loss due to SKP Ordered Sets for common clock (higher for other modes), and 64/68 represents the Flit overhead (2 Bytes each for protocol ID and Flit CRC). This results in a link efficiency of 0.924 with sync Hdr on and 0.939 when sync Hdr is off. This includes credits and link reliability mechanisms such as Ack/Nak (Ack: Acknowledge, Nak: Negative Acknowledge) [8].

For 256-Byte and 128-Byte Latency-Optimized (128B LO) Flits, for CXL.cache and CXL.mem, the overall link efficiency is 15/16 (=0.938), since there are 15 slots and one slot equivalent is spent on FEC/ CRC/ Hdr etc. [8,9]. The slots for CXL.cache and CXL.mem are similar even after accounting for the extra bits for scalability. As a result, as can be seen later, the bandwidth efficiency tends to be similar across the three Flit types.

There are three common traffic mixes to benchmark performance in CXL.io and PCIe: 100% Reads, 100% Writes, and 50-50 Read-Write. Table 3 summarizes the realized CXL.io bandwidth for the various payload sizes for each of the three traffic mixes for each of the three Flit types, based on the methodology described in [8]. Typical workloads include a mix of small payloads (4B-32B) as well as the medium to large sized payloads (64B and above). Even though commercial systems deploy either a max payload size of 128B (or 32 Double Words aka. DWs) for client CPUs and 256B/512B (64/128 DW) for server CPUs, we have provided the projections up to the maximum 4KB (1024 DW) payload size for completeness.

*Table 3: CXL.io realizable bandwidth in GB/s for different traffic mixes and payload size for the CXL 2.0 68B Flits and CXL 3.0 256B and 128B Flits for a x16 link at 32GT/s. See [8] for more details.*

| Payload Size (DW) | 68-Byte Flit | | | 256-Byte Standard Flit | | | 128B Latency-Optimized Flit | | |
|---|---|---|---|---|---|---|---|---|---|
| | 100% Read | 100% Write | 50-50 RW | 100% Read | 100% Write | 50-50 RW | 100% Read | 100% Write | 50-50 RW |
| 1 | 9.8 | 8.4 | 9.1 | 14.7 | 11.8 | 13.1 | 14.5 | 11.6 | 12.9 |
| 4 | 26.2 | 23.5 | 29.4 | 33.6 | 29.4 | 39.2 | 33.1 | 28.9 | 38.6 |
| 16 | 44.9 | 42.8 | 67.3 | 49.6 | 47.1 | 78.5 | 48.7 | 46.3 | 77.1 |
| 64 | 54.6 | 53.8 | 99.2 | 56.2 | 55.4 | 104.6 | 55.3 | 54.4 | 102.8 |
| 256 | 57.7 | 57.5 | 112.5 | 58.2 | 57.9 | 114.1 | 57.2 | 57.0 | 112.2 |
| 1024 | 58.6 | 58.5 | 116.4 | 58.7 | 58.6 | 116.8 | 57.7 | 57.6 | 114.8 |

(100% Read is outbound data flow only; 100% Write is inbound data flow only; 50-50 RW is equally distributed between the two directions. As expected, the bandwidth reaches the raw bandwidth of 64 GB/s/direction as the payload size increases due to amortization of the same overhead over a larger payload. Also, the 256B and 128B LO Flits get better bandwidth efficiency since they do not have the framing and per-TLP CRC overhead)

**CXL.cache**: A device reading a cache line from memory using RdCurr in D2H results in an H2D Data_Hdr plus the H2D Data. Each H2D Data_Hdr is 24 bits, 4 of these can be packed in a slot, representing 4 cache line transfers in a slot. Each H2D Data is 64-Bytes and hence needs 4 slots. Thus, a x16 CXL device would get (16/17) *0.94*64 GB/s = 56.6 GB/s of bandwidth with reads from the processor with 68-Byte Flit format. For 256-Byte and 128-Byte Latency-Optimized Flits, the data is the bottleneck since each H slot can have up to 4 H2D Data_Hdr. For a 256-Byte Flit, we only have 14 slots available for data, which can accommodate 3.5 cache lines. Hence the realizable bandwidth is: (14/16) *128=112 GB/s for 256-Byte Flit and (13/16) *128=104 GB/s for 128-Byte Latency-Optimized Flit [8, 9].

For writes, the device issues a D2H Req (RdOwn). This results in an H2D response with Data. The device then issues a D2H Dirty Evict, obtains an H2D Resp (Wr Pull) which causes it to do a D2H Mem Wr. This results in 3 Hdrs (D2H Req RdOwn, D2H Dirty Evict + Data Hdr) for every cache line worth of data in the D2H direction. For 68-Byte Flit, the 3 Hdrs occupy two slots (e.g., D2H Req RdOwn+ Data Hdr, D2H Dirty Evict) resulting in (4/6) *0.94*64 GB/s = 40 GB/s per direction. For the 256-Byte as well as 128-Byte Latency-Optimized Flit Formats, that is 4 G-slots and 2.25 any slots (one each for the Req and Evict Dirty and ¼ for the Data_Hdr since we can have 4 of them in any slot) in D2H direction, resulting in a realizable data bandwidth of (4 data slots/6.5 total slots)*(15/16)*128 = 73.8 GB/s [8, 9].

**CXL.mem**: A memory read from CPU is sent as MemRd in M2S (Section 3.5). It results in a response which comprises of a S2M DRS Hdr (MemData) and 4 slots of Data (64B) for a Type-3 device. If the device is a Type 2 device, an additional S2M NDR for completion (Cmp) is needed. For the writes, the CPU sends an M2S Req (MemWrite) and 4 slots of Data, generating an S2M NDR as Cmp.

An implementation will attempt to pack multiple headers in a slot to maximize efficiency. The CXL specification allows different packing permutations of various headers into a slot, which are denoted using a so-called H* notation [8].

There are a few representative workloads to evaluate the memory bandwidth performance. 100% Reads which is represented as 1R0W; 100% Writes, which is represented as 1R1W since every cache line write involves reading the content of memory before updating it; and equal read and write which represents 2 cache line reads for every write. Table 4 summarizes the realizable bandwidth for CXL.mem for different traffic mixes. The detailed derivation appeared in [8].

When there are only reads with no writes, there is no real data transfer in the M2S direction (only read requests go). Hence the data bandwidth is 0 in that direction whereas the S2M direction is mostly data (1 slot of header for 2DRS followed by 8 slots of data for the 2 cache lines). Hence the data efficiency is 0.939 for the link efficiency x 8/9 for the slot efficiency x 64 GB/s raw bandwidth = 53.5 GB/s. Other entries follow a similar approach, as detailed in [8]. The scheduler in SPR Flit packing logic follows a greedy algorithm for CXL.mem by prioritizing H3 for the first 16B slot in a Flit followed by H5 and populating data in the remaining Flits if less than 64Bytes of Data is scheduled to be sent. If 64 Bytes or higher Data is scheduled to be sent, it schedules an all-Data Flit. This approach results in the measured bandwidth very close to the numbers in Table 4 for the 68-B Flit mode.

*Table 4: Realizable bandwidth in GB/s with CXL.mem for different traffic mixes across CXL 2.0 68B Flits and CXL 3.0 256B and 128B Flits for a x16 link at 32GT/s. See [8] for more details.*

| Data B/W (GB/s)<br>x16 @ 32G or x8 @ 64 G<br>(Raw B/W: 64 GB/s/dir) | | Type-3 | | Type-2 | |
|---|---|---|---|---|---|
| | | M2S | S2M | M2S | S2M |
| 1R/0W | 68B Flit | 0 | 53.5 | 0 | 48.1 |
| | 256B Flit | 0 | 54.0 | 0 | 50.3 |
| | 128B LO | 0 | 51.9 | 0 | 49.1 |
| 1R/1W | 68B Flit | 40.1 | 40.1 | 40.1 | 40.1 |
| | 256B Flit | 39.9 | 39.9 | 39.9 | 39.9 |
| | 128B LO | 39.9 | 39.9 | 39.9 | 39.9 |
| 2R/1W | 68B Flit | 25.3 | 50.7 | 22.9 | 45.8 |
| | 256B Flit | 26.0 | 52.1 | 24.3 | 48.6 |
| | 128B LO | 25.4 | 50.9 | 24.3 | 47.5 |

For the 256-Byte standard or 128-Byte Latency-Optimized Flit, a designer needs to consider additional constraints. For example, one may use the H/ HS slot for header only. The scheduling will ensure that no G-slot will go empty when there is data (or header) to be sent. Thus, headers that precede data will be prioritized and opportunistically placed in H/HS slots while making sure that no more than 5 cache lines worth of data are to be scheduled. We also ensure that other headers make forward progress.

With the UIO/BI flows introduced in CXL 3.0 [6, 8, 9], the host processor is bypassed, and accesses go directly between devices using the CXL.io UIO flows both for HDM accesses as well as inter-domain messages that do not need caching. The expectation is that the vast majority of these will not cause invocation of the BI-Snp mechanism to enforce coherency since I/O devices and cores are typically not accessing the same data simultaneously. This helps with link efficiency as well as congestion on the host-processor links. In the analysis of Table 5, the efficiency gain with this mechanism is significant, even in the pathological case where 100% of accesses cause BI-Snp (x =1.0 cases in Table 1b). Such simultaneous accesses from the host and device may occur for control data structure that are used for synchronization. Typically, we would expect such data structures to be placed in host memory. However, even if 100% of accesses cause BI-Snp, UIO/BI are better than multiple cache line transfers across links.

*Table 5: Link Efficiency with Back-Invalidate (BI) and Unordered IO (UIO) Flows [9]*

| | In (DW) | Out (DW) | M2S (Slot) | S2M (Slot) |
|---|---|---|---|---|
| 100% Read (Existing Flow) | 5.*a* (Mem Rd X) | *(4+d).a* (Cpl X) | *b. d'* (Req X - 1 per slot) | *5.b.d'* (1 Data Hdr, 4 Data – 5 slots) |
| 100% Read (w/ BI Flow) | *5.c* (Mem Rd X) | *(4+d).c* (Cpl X) | *(x/3).c.d'* (BI Rsp – 3 per slot) | *x.c.d'* (BI Snp) |
| 100% Write (Existing Flow) | *(5+d).a* (Mem Wr X) | *4.a* (Cpl X) | *6.b.d'* (1 Rd Req, 1 Wr Req, 4 Data – 1 slot per Req) | *(4.67).b.d'* (1 DRS + 1 NDR in 2/3 slot, 4 Data for X) |
| 100% Write (w/ BI Flow) | *(5+d).c* (Mem Wr X) | *4.c* (Cpl X) | *(5.33)x.c.d'* (1 Req - 1 slot, 1 BI Rsp - 1/3 slot, 4 Data Slots) | *x.c.d'* (1 BI Snp + 1 NDR in 1 slot) |

A. Number of bytes required for device accesses where α= Hops between device and CPU, b = Hops between CPU and memory, and c = Hops between device and memory, d = payload in Double Words (DWs), d': No. of cache lines for d DWs = ceil (d/16), x = Ratio of accesses that requires a BI -Snp.

| Data Size (DW) | 1 | 4 | 8 | 16 | 24 | 32 | 64 | 128 |
|---|---|---|---|---|---|---|---|---|
| 100% Read (a=b=c=2, x= 0.1) | 3.17 | 2.69 | 2.30 | 1.89 | 2.34 | 2.08 | 2.21 | 2.29 |
| 100% Write (a=b=c=2, x= 0.1) | 4.13 | 3.52 | 3.00 | 2.42 | 3.06 | 2.70 | 2.88 | 2.99 |
| 100% Read (a=b=c=2, x= 1.0) | 1.24 | 1.20 | 1.16 | 1.12 | 1.16 | 1.14 | 1.15 | 1.16 |
| 100% Write (a=b=c=2, x= 1.0) | 1.49 | 1.45 | 1.41 | 1.34 | 1.41 | 1.37 | 1.39 | 1.40 |

Detailed explanation: a Memory Read transaction in CXL.io is 5DW long and takes *a* hops to reach the CPU with existing flows, accounting for 5*a*; the resulting completion in CXL.io is 4DW hdr plus *d* DW data, accounting for *(4+d)a* in Row 1. This read causes *d'* cache line requests in CXL.Mem from the CPU to the CXL device, which are reflected in the M2S and S2M directions in Row 1. With the proposed BI flow, the CXL.io requests go directly to the device with *x* fraction of cache lines causing BI to CPU in CXL.Mem, as reflected in Row 2 entries. We add the values of each Row – the slots are multiplied by 4 to convert to DWs. CXL.io traffic assumes 10% FEC/CRC/DLLP overhead. All M2S/S2M Slots are assumed to be 16Bytes for simplicity and use 1/15 FEC/CRC/HDR overheads. We get similar results for a=b=c= 1 or 2 or 3.

# 7 DISCUSSION

We discuss four implications for the impact of CXL on the compute landscape and then outline an initial set of future directions.

## 7.1 Implications for the compute landscape

**Implication 1: CXL network effects for devices**. Compared to PCIe, CXL significantly expands usage models and capabilities, e.g., due to lower latency, pooling, and hardware coherency. These advantages create incentives for I/O devices to adopt CXL. Initially, this will likely apply most to accelerators and memory/storage devices, where CXL.cache and CXL.mem offer massive performance uplift and CXL pooling offers TCO savings. Eventually, CXL adoption will likely create a network effect where other IO-devices like network interfaces must offer CXL because CXL will have become the common access mechanism. Near-memory or in-memory processing systems are similarly likely to adopt CXL due to its significantly simpler programming model enabled by its strong support for cache coherency.

**Implication 2: migration of memory to parallel bus on package and CXL off-package**. Compared to locally attached memory (DDR), CXL offers fundamental scaling and efficiency advantages. CXL offers 8x higher bandwidth per pin and higher capacity, which gives CXL a significant scalability advantage. Memory pooling and heterogeneous memory media give CXL a cost advantage [58]. CXL also helps thermals as it enables longer and more flexible motherboard routing compared to DDR. These advantages come at a higher idle latency than DDR (often twice the latency). When considering loaded latency, CXL may perform better than DDR due to its bandwidth advantage and on-package memory would significantly outperform either. A natural expectation is that more memory will migrate to CXL and on-package memory. Eventually CXL will become the only external memory attach point for CPUs and accelerators.

**Implication 3: CXL will grow to become a rack or cluster-level interconnect.** Compared to today's datacenter networks based on Ethernet and InfiniBand, CXL lowers latency by an order of magnitude. Additionally, CXL's coherent memory sharing and fine-grained synchronization can significantly boost distributed system performance for key workloads such as large machine learning models and databases. However, CXL requires dedicated cabling with stringent requirements, e.g., on cable length, adaptors, and use of retimers. This leads to significantly higher cost and less flexibility than today's datacenter networks. It is thus likely that CXL will be deployed within a smaller domain, such as within racks or across a few racks (cluster or pod). For example, current financial models point to sub-rack CXL deployments as the TCO sweet spot [58]. With improved cost through standardization, CXL's scope can likely grow, but it is unlikely that CXL will replace Ethernet as the datacenter-wide networking standard.

**Implication 4: CXL will enable highly composable systems.** Composability means that components and resources can be dynamically assembled at run time and assigned to a workload or virtual machine. We expect that both memory devices and IO devices (e.g., NICs, accelerators, and storage) will evolve multi-host capabilities that allows dynamically assigning fractions of their capacity to individual hosts over CXL. Converged and pooled devices lead to significantly better resource usage due to increased multiplexing opportunities, and thus lower cost. This design also facilitates accelerating distributed systems via shared memory, message passing, and peer-to-peer communication via CXL. However, a composable system is still distinct from treating all resources connected via CXL as a single big system. Workloads will still prefer locality and spanning as few CXL links as possible in order to minimize coherence traffic and the blast radius of any failure.

### 7.2 Future Directions

CXL opens research directions across computer science and engineering. In computer architecture, CXL opens a path to prototype and deploy new memory controller features [61] such as adaptive DRAM refresh to save power and improve performance [62, 63], low-cost improvements of reliability [64], and reducing memory waste [65, 66]. External memory controllers can evolve independently from the main CPU and are lower cost, which facilitates faster iteration and customization. Research may also explore implications for CPU architecture. For example, with higher memory latency and more cache lines in flight, one may expect different prefetching and buffer-sizing strategies as well as increased interest in reducing memory access latencies [69, 70]. Another challenge might be impacts from delayed hits which require more rigorous simulations and modeling [60] as well as different cache management strategies [59].

In computer systems, CXL memory pooling will spark changes to both local and distributed memory management. For example, we need mechanisms that can prioritize workloads and guarantee performance in the face of memory pressure from remote hosts. We may also need systems-level approaches to mitigate congestion as well as failures in a distributed memory fabric. For example, QoS in the CXL standard is currently limited to CXL.mem and does not address fabric congestion. Workload scheduling will also need to evolve to support composability, with a significantly enlarged design space and different definitions of locality compared to existing datacenter designs. The CXL specification needs to evolve to provide QoS across other protocols while handling fabric congestion and supporting dynamic multi-pathing at a transaction level for QoS as well as for fail-over.

In computer engineering, research and development needs to continue to reduce the latency of CXL memory, accelerators, and switches to expand the applicability of CXL. There are many opportunities including iterating on engineering CXL building blocks, exploiting process advancements (Section 6.1), and better packing algorithms (Section 6.3). The ecosystem also needs to develop a rigorous approach to error containment and management of CXL's increased blast radius. Working around faults and congestion hot spots with load-store semantics will have different constraints than networking applications which can handle lost packets as well as completely out-of-order delivery of packets. Finally, CXL can be further enhanced and deployed to expand across multiple racks and offer high reliability with low-latency load-store access for multiple applications with dynamic fail-over and finer-grained quality of service (QoS) enhancements that will be incorporated in future revisions of the specification. With co-packaged optics enabled through Universal Chiplet Interconnect Express (UCIe) retimers [10, 50, 74, 75, 76], we expect to realize the vision of building composable and scale-out systems spanning the rack through the pod at the datacenter, resulting in power efficient performance with significant total cost-of-ownership benefits.

### 7.3 Conclusions

CXL addresses significant industry challenges while following a design philosophy based on openness, simplicity, and backward-compatibility. This has enabled CXL to gain wide traction as a common standard across the industry. All of these factors make CXL a trending research area within the academic community as well. We hope that this tutorial serves as both an introduction and jumping off point into the standard as well as a base for research ideas.